\documentclass[journal,draftcls,onecolumn,12pt]{IEEEtran}

\usepackage[utf8x]{inputenc}
\usepackage[T1]{fontenc}
\usepackage{eurosym}
\usepackage{color}
\usepackage{bbold}
\usepackage{graphicx}
\usepackage[parfill]{parskip} 
\usepackage{array}
\usepackage{amsmath}
\usepackage{subfigure}
\usepackage{stmaryrd}

\usepackage{cite}
\usepackage{url}

\usepackage{bm}
\usepackage{bbm}

\usepackage{macro}

\usepackage[ruled,vlined]{algorithm2e}

\title{Optimizing the Energy Efficiency of Unreliable Memories for Quantized Kalman Filtering}

\author{Jonathan Kern\textsuperscript{1,3}, Elsa Dupraz\textsuperscript{1}, Abdeldjalil Aïssa-El-Bey\textsuperscript{1}, Lav R. Varshney\textsuperscript{2}, François Leduc-Primeau\textsuperscript{3} \\
\thanks{Part of the material of this paper was published in ICASSP 2021. This work was supported by grant ANR-17-CE40-0020 (EF-FECtive project) and by the ``Make our Planet Great Again'' Initiative of the Thomas Jefferson Fund. }
\small \textsuperscript{1} IMT Atlantique, Lab-STICC, UMR CNRS 6285, F-29238, France \\
\textsuperscript{2} Coordinated Science Laboratory, University of Illinois at Urbana-Champaign, USA \\
        \textsuperscript{3} Department of Electrical Engineering, École Polytechnique de Montréal, Montreal (QC), Canada}

\begin{document}

\maketitle
\thispagestyle{empty}

\begin{abstract}
This paper presents a quantized Kalman filter implemented using unreliable memories. We consider that both the quantization and the unreliable memories introduce errors in the computations, and develop an error propagation model that takes into account these two sources of errors. In addition to providing updated Kalman filter equations, the proposed error model accurately predicts the covariance of the estimation error and gives a relation between the performance of the filter and its energy consumption, depending on the noise level in the memories. Then, since memories are responsible for a large part of the energy consumption of embedded systems, optimization methods are introduced so as to minimize the memory energy consumption under a desired estimation performance of the filter. The first method computes the optimal energy levels allocated to each memory bank individually, and the second one optimizes the energy allocation per groups of memory banks. Simulations show a close match between the theoretical analysis and experimental results. Furthermore, they demonstrate an important reduction in energy consumption of more than 50\%.
\end{abstract}
\section{Introduction}
\IEEEPARstart{K}{alman} filtering is a very common recursive estimation task in statistical signal processing~\cite{kalman_new_1960}, and it is often implemented on resource-limited hardware. Applications which require an embedded energy-efficient Kalman filter include air quality monitoring~\cite{lai_iot_2019}, biomedical wearable sensors~\cite{anania_development_2008}, and vehicle positioning~\cite{sung_simplified_2020}. Energy budgets for embedded systems show that memory access consumes about a hundred times more energy than integer computations~\cite{horowitz_11_2014}. Therefore, in this paper, we focus on optimizing the energy used by memories in Kalman filters.
 
All memories used in integrated circuits exhibit a fundamental trade-off between data storage reliability and energy consumption that is related to the inability of perfectly controlling the fabrication process. For example, the energy consumption of static random access memories (SRAMs) can be reduced by lowering its supply voltage, but this increases the probability that some of the stored bits cannot be retrieved correctly~\cite{dreslinski_near-threshold_2010}.
Following this principle,~\cite{kim_generalized_2018} developed an optimization method to lower the energy consumed by SRAM accesses by reducing the bit-line voltages. This methodology was also used to decrease the write energy of magnetic random access memories (MRAM) in~\cite{kim_optimizing_2020}.
In both cases, however, this introduces errors in the words stored in memory. 

The robustness to unreliability in computation operations and memories has been investigated for several signal processing and machine learning applications, including binary recursive estimation~\cite{dupraz_binary_2019}, binary linear transformation~\cite{yang_computing_2017}, deep neural networks~\cite{henwood_layerwise_2020,hacene_training_2019} ,or distributed logistic regression~\cite{yang_fault-tolerant_2016}.
Moreover, several techniques have been proposed to compensate for faults introduced by unreliable systems. For instance,~\cite{hegde_energy-efficient_1999} proposed to add redundancy in the system through algorithmic noise tolerance, and~\cite{huang_acoco_2015} investigated the use of error-correction codes (ECC) for fault correction. 

Although Kalman filtering has not previously been investigated under unreliable hardware implementation, some related works considered inaccuracies in the filter parameters~\cite{huang_novel_2018,yang_robust_2001,hounkpevi_robust_2007},  and uncertainties on the observations~\cite{nahi_optimal_1969}. Although these models are not relevant for characterizing the effect of unreliable memories, the main lessons they provide is that Kalman filtering is very sensitive to inaccuracies, and that one should re-derive the optimal Kalman filter depending on the specifically considered uncertainty model. 
Other prior work aimed at reducing energy requirements for Kalman filtering focus on reduced computational complexity in FPGA~\cite{jarrah_optimized_2016,sunil_kumar_optimization_2020} and ASIC~\cite{pereira_exploring_2019} implementations.


Designing a digital hardware implementation requires quantizing all the variables and computational operations. 
Therefore, to further reduce the memory energy consumption, one option is to properly optimize the quantization so as to reduce the memory requirements of the implementation. 
Significant energy gains from optimized quantization have been demonstrated in~\cite{wang_reducing_2015,xia_energy-efficient_2019,marcastel_energy_2019} for signal processing and digital communications applications, and in~\cite{hashemi_understanding_2017,ding_quantized_2018,jain_compensated-dnn_2018} for neural networks. 
The effects of quantization on the Kalman filter were first studied in~\cite{stripad_performance_1981,verhaegen_numerical_1986}  to understand the convergence of filters with reduced precision. 
More recently,~\cite{sun_quantized_2007,li_distributed_2015,hu_quantized_2018} considered two distributed quantized Kalman filters, one based on quantized observations and one based on quantized innovations, where sensors process and transmit quantized observations and innovations to a fusion center. 
Furthermore,~\cite{sun_quantized_2007} proposed to optimize the number of quantization bits at each sensor so as to minimize the required data transmission energy.
More general linear stochastic systems were also investigated under quantized measurements~\cite{you_quantized_2011} and quantized innovations~\cite{you_recursive_2009}, where it was shown that the derived quantized filters converge to standard Kalman filters as the number of quantization levels increases.
However, none of these theoretical works considered quantized parameters (\emph{\text{e.g.}}, quantized Kalman gain matrices, quantized measurement matrices, etc.), in addition to quantized observations/innovations. 
Therefore in this paper, we study a fully-quantized Kalman filter and investigate its energy consumption when using unreliable memories. 


Here, we aim to optimize the energy consumption of a Kalman filter implemented with fixed-point quantization~\cite{dally_digital_2015} and with unreliable memories. We consider the statistical model of~\cite{dreslinski_near-threshold_2010} which relates the amount of faults introduced in memory to its energy consumption. Then, as a first contribution, we propose a unified framework to analyze the performance of Kalman filters with both quantization errors and faults introduced in the memory. To develop this framework, we build on the approach of~\cite{verhaegen_numerical_1986} which consists of evaluating the covariance matrix of the estimation error at each filter iteration, by considering both error propagation from previous iterations, and errors introduced at the current iteration.  
Our analysis also includes quantized filter parameters, and further incorporates the effect of unreliable memories. Note that determining the covariance matrix of the estimation error has two advantages. First, it allows us to derive the optimal Kalman filter equations under the considered quantization and memory error models. Second, and more specific to our case, it defines a performance criterion which will be used to optimize the memory energy consumption. 

As a second contribution, we define two optimization problems to minimize the memory energy consumption while satisfying a target constraint on the estimation performance of the Kalman filter.  
In the first problem, we optimize the number $B$ of quantization bits and the energy allocated to each bit position so as to minimize the overall energy consumption of the memory. This optimization problem extends~\cite{kim_generalized_2018}, which was not dedicated to Kalman filtering but derived optimal bitwise energy allocations with a fixed number of quantization bits, by considering a generic Mean-Squared Error (MSE) performance criterion when reading a word in memory. 
Although a useful baseline, the setting where each bit position can have a different energy allocation is not practical, since each of the $B$ bits should be placed in a different memory bank with its own power supply. This is why we also introduce a second optimization problem in which we fix the number $L < B$ of possible energy levels, and optimize the energy value in each level and the mapping of bit positions to an energy level.
At the price of a small energy increase, this optimization problem allows us to build a practical implementation which only requires $L$ memory banks. 
By using the Karush-Kuhn-Tucker (KKT) conditions, we provide solutions for the two considered optimization problems. Both solutions can be numerically computed using water-filling. Numerical simulations show that after optimization, the memory energy consumption is reduced by up to $56\%$ compared to uniform allocation.

The main contributions of this paper can be summarized as:
\begin{enumerate}
    \item We develop an error propagation model of the Kalman filter that takes different sources of errors (quantization, unreliable memories) into account and allows us to derive new filter equations to minimize the estimation error. Moreover, these new equations accurately predict the filter's performance, depending on the considered sources of errors and on their parameters.
    \item We propose a methodology for minimizing the energy of the unreliable memories used in the Kalman filter, under a given performance constraint. 
    It consists of computing the optimal number of quantization levels and bit energy allocation in two setups. The first setup considers that the $B$ energy levels can be chosen freely, while the second one assumes that only $L<B$ energy levels can be set.
\end{enumerate}

A preliminary version of this paper \cite{kern_improving_2021} only considered optimizing the energy allocation for each memory bank individually and for a fixed number of bits, without taking into account the quantization noise nor trying to reduce the energy consumption by adjusting the number of bits. 

The rest of the paper is organized as follows.  Section~\ref{sec:kalman} describes the quantized Kalman filter and introduces the uncertainty model for unreliable memories.  Section~\ref{sec:error} investigates the theoretical performance of the filter. Section~\ref{sec:opti} formally defines and solves the two considered optimization problems. Section~\ref{sec:simu} presents simulations results.

\section{System Model}\label{sec:kalman}
We first review the Kalman filter for estimating dynamic state variables from noisy measurements. We then present the considered implementation of the filter, by first introducing its quantization model and then describing its implementation with an unreliable memory.

\subsection{Kalman filter}

Consider a linear dynamic variable $\bm{x} \in \mathbb{R}^c$  described by a process:
\begin{equation}
    \bm{x}_{k+1} = \bm{F}\bm{x}_{k}+\bm{u}_k  \,,
\end{equation}
where $\bm{x}_{k}$ is the state vector of the process at step $k$, $\bm{F}$ is the $c \times c $ state transition matrix, and $\bm{u}_k \in \mathbb{R}^c$ is an additive white noise vector.
The state of $\bm{x}$ is observed through the measurement vector $\bm{y} \in \mathbb{R}^d$:
\begin{equation}
    \bm{y}_k = \bm{H}\bm{x}_k+\bm{v}_k \,,
\end{equation}
where $\bm{H}$ is the $d \times c$ measurement model and $\bm{v}_k \in \mathbb{R}^d$ is an additive white noise on the measurements, independent from the model noise $\bm{u}_k$.
The covariance matrices of the noise vectors $\bm{u}_k$ and $\bm{v}_k$ are known and denoted $\bm{Q}$ and $\bm{R}$, respectively.

The Kalman filter recursively estimates the successive states $\bm{x}_k$ from the measurement vectors $\bm{y}_k$ and from the known model, by minimizing the mean squared error $\operatorname{MSE}(\bm{x}) = \mathbb{E}[\lVert \bm{x}_k- \bm{\hat{x}}_{k}\rVert^2]$ between $\bm{x}_k$ and its estimate $\bm{\hat{x}}_{k}$ at each step $k$. 
The filter can be decomposed in two phases: the a priori estimation uses only the known model, and the a posteriori estimation takes into account the measurements.
Each phase computes both estimates $\bm{\hat{x}}_{k+1|k}$ (a priori phase) and $\bm{\hat{x}}_{k+1|k+1}$ (a posteriori phase) of the state vector $\bm{x}_{k+1}$, and the covariance matrices of the estimations errors $\bm{P}_{k+1|k} = \mathrm{Cov}[\bm{x}_{k+1}-\bm{\hat{x}}_{k+1|k}]$ and $\bm{P}_{k+1|k+1} = \mathrm{Cov}[\bm{x}_{k+1}-\bm{\hat{x}}_{k+1|k+1}]$.
The recursive equations of the a priori estimation step are: 
\begin{align}
    & \bm{\hat{x}}_{k+1|k} = \bm{F}\bm{\hat{x}}_{k|k} \label{eq:priori_x} \,, \\
    & \bm{P}_{k+1|k} = \bm{F}\bm{P}_{k|k}\bm{F}^\top+\bm{Q} \label{eq:P_priori} \,,
\end{align}
and the recursive equations of the a posteriori estimation step are: 
\begin{align}
    &  \bm{K}_{k+1} = \bm{P}_{k+1|k}\bm{H}^\top(\bm{H}\bm{P}_{k+1|k}\bm{H}^\top+\bm{R})^{-1} \,, \\
    &  \bm{\hat{x}}_{k+1|k+1} = \bm{\hat{x}}_{k+1|k}+\bm{K}_{k+1}(\bm{y}_{k+1}-\bm{H}\bm{\hat{x}}_{k+1|k}) \label{eq:posteriori_x} \,, \\
    &  \bm{P}_{k+1|k+1} = (\bm{I}-\bm{K}_{k+1}\bm{H})\bm{P}_{k+1|k} \label{eq:P_posteriori}  \,,
\end{align}
where $\bm{A}^\top$ denotes the transpose of a matrix $\bm{A}$.
In these equations, the covariance matrices $\bm{P}$ of size $c \times c$ and the Kalman gain $\bm{K}$ of size $c \times d$ can be computed offline. On the other hand, the terms $\bm{\hat{x}}_{k+1|k}$ and $\bm{\hat{x}}_{k+1|k+1}$ depend on the measurements $\bm{y}_{k}$ and must be computed online.

\subsection{Quantized implementation of the filter}
\label{sec:quanti}

In the rest of the paper we study Kalman filters that are implemented under fixed-point quantization~\cite{dally_digital_2015}. 
Under this model, each number is represented as a signed integer coded on $(1+n+m)$ bits, where one bit is used for the sign, $n$ bits are used for the integral part of the number, and $m$ bits are used for its fractional part. Using this model, a given number $z$ can be written as 
\begin{equation}\label{eq:bin_rep}
z= (-1)^{z_n}\sum_{b=-m}^{n-1}2^b z_b,
\end{equation}
where $z_b \in \{0,1\}$ are the bits stored in memory to represent $z$. In our modeling of the Kalman filter, all variables (including matrix components) involved in equations~\eqref{eq:priori_x}--\eqref{eq:P_posteriori} are stored using this quantization model, all with the same values of $n$ and $m$. The quantization of the variables to this fixed-point model is done using a uniform quantizer. Note that the distribution of the quantized data is not necessarily uniform (the random variables $\hat{\bm{x}}_{k|k}$ and $\bm{y}_k$ could follow Gaussian distributions for example). However, in~\cite{ziv_universal_1985} it is shown that a uniform quantizer can be applied independently of the probability distribution of the source with only a small difference to an optimal quantizer. 

In the considered quantizer, the value of $n$ is chosen so as to be able to represent the largest possible value in the system. The value of $m$ sets the resolution of the quantization so the smallest difference between two quantized numbers is $2^{-m}$~\cite{dally_digital_2015}. The value of $m$ will be a parameter that is optimized for minimizing the energy in later sections.

In the case of fixed values such as components of the matrices of the filter, the fixed-point quantized value can be written as $\Q{f} = f + \delta_f$ where $\delta_f$ is the quantization error. Using the previously described uniform quantizer, $\delta_f<2^{-m}$.
In the case of quantized random variables like the components of $\bm{\hat{x}_{k|k}}$ or $\bm{y_{k}}$, we let $\epsilon_x$ be the quantization error, and express $\Q{x} = x+\epsilon_x$. In~\cite{sripad_necessary_1977}, conditions are given for the quantization error $\epsilon_x$ to be independent from the quantized variable, depending on the distribution of the quantized data. For the special case of a Gaussian distribution, the quantization step needs to be significantly smaller than the variance of the quantized data. In this case, it can be shown that the quantization error is a white noise following a uniform distribution of variance $\frac{2^{-2m}}{12}$. This independence assumption will be used in the theoretical derivations of Section~\ref{sec:error}.
Note that most existing works on quantized Kalman filters only consider that random quantities like $\bm{\hat{x}_{k|k}}$ and $\bm{y_{k}}$ as quantized, whereas here, the components of the matrices \emph{e.g.}, $ \bm{K}_{k}$, of the filter are also quantized. This will require a new theoretical analysis to treat this case.


\subsection{Implementation of the filter by using an unreliable memory}
\label{sec:unreliable_Kalman}

In order to reduce its energy consumption, the quantized Kalman filter can be implemented on unreliable hardware~\cite{kim_generalized_2018,yang_computing_2017,dupraz_binary_2019,henwood_layerwise_2020}. Here, we assume, as in~\cite{dupraz_binary_2019} and~\cite{henwood_layerwise_2020}, that only the memory is faulty. In this case, each memory cell of a memory bank has a bit flipping probability $p$. We then use the model of~\cite{dreslinski_near-threshold_2010} to express $p$ with respect to the memory bank energy consumption $e$ as
\begin{equation}
\label{eq:link_proba_energy}
    p = \exp(-e a) \,,
\end{equation}
where $a$ is a parameter that depends on the device technology. We assume that the bit errors from each memory cell are independent. 

Each memory bank has a uniform energy consumption (e.g. single supply voltage) and is used in our case to store the bits at a certain position of all components of matrices that are stored in the unreliable memory. 
Here we consider that only the estimates $\bm{\hat{x}}_{k+1|k}$ and $\bm{\hat{x}}_{k+1|k+1}$ are stored in unreliable memory bank. We make this assumption since the other terms of the filter can be precomputed offline and stored on a reliable memory separately in the system. Therefore, in the Kalman filter, instead of having an estimate component $\hat{x}$, such as one the computed in~\eqref{eq:priori_x}, we have a possibly incorrect estimate component $\tilde{x}$. Using the binary representation given in~\eqref{eq:bin_rep}, we define an energy per memory bank vector:
\begin{equation}
\label{eq:energy_vector}
\bm{e} = \begin{bmatrix}e_{-m} , e_{-m+1} , \ldots , e_{n-1}\end{bmatrix} \,.
\end{equation}
A bit at position $b$ stored in the unreliable memory can then be expressed as $\tilde{x}_b = \hat{x}_b \oplus \gamma_b$,
where $ p_b =  \Pr(\gamma_b = 1) =  \exp(-e_ba)$, and $\oplus$ denotes the modulo-2 addition.
As the filter would be particularly sensitive to faults on the sign bit, we consider a sign-preserving model, as in~\cite{dupraz_analysis_2015,dupraz_binary_2019,ngassa2015density}. This sign-preserving model can be implemented by storing the sign bits in a separate reliable memory.

Using this noise model defined at the bit-level $\tilde{x}_b$, we can define a noise model at the symbol level $\tilde{x}$ as
\begin{equation}
    \tilde{x} = \hat{x}+\gamma \,,
\end{equation} 
where $\gamma$ is the noise introduced by the unreliable memory.
For the subsequent theoretical analysis, we assume that the mean $\mathbb{E}[\gamma]$ of this memory noise is negligible compared to its variance $\operatorname{Var}[\gamma] = \sigma_\gamma^2$. We verified this condition from Monte Carlo simulations.
The covariance matrix $\bm{\Gamma}$ of a memory noise vector $\bm{\gamma}$ of length $c$ is defined as $\bm{\Gamma} = \mathrm{Cov}[\bm{\gamma}] =     \bm{I}_c \sigma_\gamma^2 \,,$
and has size $c \times c$. The matrix $\bm{\Gamma}$ is diagonal since the memory noise variables are considered independent.

\section{Error analysis}\label{sec:error}
As described in Section~\ref{sec:kalman}, we consider two types of errors affecting the filter: the quantization error and the unreliable memory noise. In this section, we first describe a generic model of error propagation in the Kalman filter, before studying both types of errors in more detail. Finally, we compute the new covariance matrix $\bm{P_{k|k}^*}=\cov{\bm{\tilde{x}}_{k|k}-\bm{x}_{k}}$ of the total estimation error by taking both sources of noise (quantization and unreliable memories) into account, compared to a standard Kalman filter which does not include either. 

\subsection{Error propagation model}
\label{sec:propag}
Our objective is to compute the total error $\bm{\Delta \hat{x}_{k+1|k+1}}$ on the computation of $\bm{\hat{x}}_{k+1}$ at step $k+1$, by considering the two types of errors: quantization and unreliable memory. 
%
To handle recursion as in~\cite{verhaegen_numerical_1986}, we choose to split the error model in two parts: the errors occurring at step $k$, and the errors from the previous steps which are propagated up to step $k$.  

To compute $\bm{\Delta \hat{x}_{k+1|k+1}}$, we first need to express the total error $\bm{\Delta P}_{k+1|k}$ on the a posteriori covariance matrix $\bm{P}_{k+1|k}$ after step $k+1$. As in~\cite{verhaegen_numerical_1986}, we express this total error as
\begin{equation}\label{eq:propag_P}
    \bm{\Delta P_{k+1|k}} = f_P(\bm{\Delta P_{k|k-1}}) + \bm{\delta P_{k+1|k}} \,,
\end{equation}
where the function  $f_P$ models the errors propagated from step $k$, and $\bm{\delta P_{k+1|k}}$ represents the errors occurring at step $k+1$.
%
In this case, according to~\cite{verhaegen_numerical_1986}:
\begin{equation}
    f_P(\bm{\delta P_{k+1|k}}) \approx  \bm{G}_{k}\bm{\delta P_{k|k-1}}\bm{G_{k}^\top} \,,
\end{equation}
where $\bm{G}_{k}=\bm{F}(\bm{I}-\bm{K}_{k}\bm{H})$.

We then express the total error $\bm{\Delta \hat{x}_{k+1|k+1}}$ on $\bm{\hat{x}_{k+1|k+1}}$ by considering the same separation between propagation errors and errors from current iteration. 
This gives
\begin{equation}
\label{eq:propag_x}
    \bm{\Delta \hat{x}_{k+1|k+1}}= f_x(\bm{\Delta \hat{x}_{k|k}},\bm{\Delta P_{k|k-1}})+ \bm{\delta {\hat{x}_{k+1|k+1}}} \,,
\end{equation}
where the error propagation function $f_x$ is provided in~\cite{verhaegen_numerical_1986} as
\begin{multline}
    f_x(\bm{\Delta{\hat{x}_{k|k}}},\bm{\Delta{P_{k|k-1}}}) = (\bm{I}-\bm{K}_{k} \bm{H})(\bm{F}\bm{\Delta \hat{x}_{k|k}}  
    +\bm{\Delta{P_{k|k-1}}}\bm{H}^\top (\bm{H}\bm{P}_{k,k-1}\bm{H}^\top+\bm{R})^{-1}(\bm{y}_{k+1}-\bm{H}\bm{F}\bm{\hat{x}}_{k,k})) \,. 
\end{multline}
In this expression, we observe the error propagation from the previous computations of $\bm{\hat{x}}_{k|k}$  $\bm{P}_{k+1|k}$. In particular,   $\bm{\hat{x}}_{k+1|k+1}$ depends on $\bm{K}_{k+1}$ which is precomputed from $\bm{P}_{k+1|k}$ at each iteration.  

Using the recursive equations~\eqref{eq:propag_P} and~\eqref{eq:propag_x}, we now estimate the covariance matrix of the total estimation error $\bm{P_{k|k}^*}$. Note that~\cite{verhaegen_numerical_1986} considered only quantization errors, while here we consider two sources of errors: quantization and unreliable memories. 
To evaluate $\bm{P_{k|k}^*}$, we must first compute the covariance of each term of $\bm{\Delta \hat{x}_{k+1|k+1}}$. By assuming that the two sources of noise (quantization and unreliable memory noise) are statistically independent, we decompose $\bm{\delta {\hat{x}_{k+1|k+1}}}$ as 
\begin{equation}
\label{eq:delta_x}
\bm{\delta {\hat{x}_{k+1|k+1}}}=\bm{\delta \hat{x}_{{k+1|k+1}}^\quant}+\bm{\delta \hat{x}_{{k+1|k+1}}^\mem}  \,,
\end{equation}
and study the two terms $\bm{\delta \hat{x}_{{k+1|k+1}}^\quant}$ and $\bm{\delta \hat{x}_{k+1|k+1}^\mem}$ separately.

\subsection{Quantization error}

We now aim for an analytical expression for $\bm{\delta \hat{x}_{k+1|k+1} ^\quant}$, defined as the difference between the full precision estimate $\bm{\hat{x}}_{k+1|k+1}$ and its quantized version $\bm{\Q{\hat{x}}}_{k+1|k+1}$:
\begin{equation}
\label{eq:delta_quanti_def}
\bm{\delta \hat{x}_{k+1|k+1}^\quant} = \bm{\hat{x}}_{k+1|k+1}-\bm{\Q{\hat{x}}}_{k+1|k+1}     \,.
\end{equation}

Before expressing  $\bm{\delta \hat{x}_{k+1|k+1}^\quant}$, we first review generic quantization errors expressions~\cite{verhaegen_numerical_1986}.
For the scalar fixed-point multiplication of a coefficient $\Q{s}$ with a random variable $\Q{t}$ both quantized according to the model presented in Section~\ref{sec:quanti}, we can show that
\begin{align}\notag
    \Q{s} \Q{t} &= (s+\delta_s)(t+\epsilon_t)+\epsilon_{st}  \\ 
    &= st+s\epsilon_t+t\delta_s+\delta_s\epsilon_t+\epsilon_{st} \label{eq:scalar_error_fp_op}\,,
\end{align}
where $\delta_s=\Q{s}-s$ and $\epsilon_t$ and $\epsilon_{st}$ follow uniform distributions of variance $\frac{2^{-2m}}{12}$.
The scalar expression~\eqref{eq:scalar_error_fp_op} can then be generalized to the case of a product between a matrix of fixed-point coefficients $\bm{\Q{A}}$ of size $p \times q$, and a matrix of fixed-point random variables $\bm{\Q{B}}$ of size $q \times r$ as
\begin{equation}
    \label{eq:quanti_dot}
    \bm{\Q{A}} \bm{\Q{B}} = \bm{A}\bm{B}+\bm{A}\bm{\epsilon_B}+\bm{B}\bm{\delta_A}+\bm{\delta_A}\bm{\epsilon_B}+\bm{\epsilon_{AB}} \,,
\end{equation}
where $\bm{\epsilon_{AB}}$ is of size $p \times r$ with ${\epsilon_{AB}}_{i,j} = \sum_{k=1}^q {\epsilon_{AB}}_{i,j,k}$. According to Section~\ref{sec:quanti}, each ${\epsilon_{AB}}_{i,j,k}$ follows a uniform distribution of variance $\frac{2^{-2m}}{12}$.
%
In~\eqref{eq:quanti_dot}, the product $\bm{\delta_A}\bm{\epsilon_B}$ can be considered as negligible compared to the other error terms. Indeed, all scalar quantization errors $\epsilon$ and $\delta$ are upper-bounded by $2^{-m-1}$ and since $m\geq 1$ their product is bounded by $2^{-2m-2}$. Thus, given that the value of $m$ is large enough, we have that  $2^{-2m-2} \ll  2^{-m-1}$. Therefore, in the following derivation, we neglect the products of quantization errors.

We now study quantization errors introduced during the computation of $\Q{\bm{\hat{x}}}_{k+1|k+1}$. While existing works, \emph{e.g.},~\cite{sun_quantized_2007,li_distributed_2015,hu_quantized_2018}, assume that only the random quantities $\bm{\Q{\hat{x}}}_{k|k}$ and $\bm{\Q{y}}_{k+1}$ are quantized, we here also consider that the matrices $\bm{\Q{D}}_{k+1}$ and $\bm{\Q{K}}_{k+1}$ are quantized as well. This corresponds to a more practical implementation setup, and requires a more complex theoretical analysis.  We first note that equation~\eqref{eq:posteriori_x} can be rewritten as
\begin{align}
    \bm{\Q{\hat{x}}}_{k+1|k+1} =\bm{\Q{D}}_{k+1}\bm{\Q{\hat{x}}}_{k|k}+\bm{\Q{K}}_{k+1}\bm{\Q{y}}_{k+1} \,,
\end{align}
where both $\bm{D}_k = (\bm{I}-\bm{K}_k\bm{H})\bm{F}$ and the Kalman gains $\bm{K}_k$ can be computed offline. We thus consider that the matrices $\bm{K}_k$ and $\bm{D}_k$ are computed in full precision and then quantized with a fixed point model.
Under these conditions, according to~\eqref{eq:quanti_dot} and neglecting the products between quantization errors, the quantized vector $\bm{\Q{\hat{x}}}_{k+1|k+1}$ can be approximated as
\begin{align}
\begin{split}
\label{eq:quant_x_approx}
            \bm{\Q{\hat{x}}}_{k+1|k+1} &\approx \bm{D}_{k+1}\bm{\hat{x}}_{k|k}+\bm{\delta_{D_{k+1}}}\bm{\hat{x}}_{k|k}+\bm{D}_{k+1}\bm{\epsilon}_{x_{k|k}} \\
            &+\bm{\epsilon}_{\bm{D}_{k+1}\bm{x}_{k|k}} +\bm{{K}}_{k+1}\bm{{y}}_{k+1}+\bm{\delta_{K_{k+1}}}\bm{{y}}_{k+1} \\
            &+\bm{K}_{k+1}\bm{\epsilon_{y_{k+1}}}+\bm{\epsilon}_{\bm{\Q{K}}_{k+1}\bm{\Q{y}}_{k+1}} \,.
\end{split}
\end{align}
We see that the expression of $\bm{\Q{\hat{x}}}_{k+1|k+1}$ depends on the full precision vectors $\bm{\hat{x}}_{k|k}$ and $\bm{y}_{k+1}$ and on the quantization errors and noise.

Finally, the quantization error $\bm{\delta \hat{x}_{k+1|k+1}^\quant}$ defined in~\eqref{eq:delta_quanti_def} can be computed by using~\eqref{eq:quant_x_approx}:
\begin{equation}
\label{eq:delta_quanti}
\begin{split}
    \bm{\delta \hat{x}_{k+1|k+1}^\quant} \approx \bm{\delta_{D_{k+1}}}\bm{\hat{x}}_{k|k}+\bm{D}_{k+1}\bm{\epsilon}_{x_{k|k}}+\bm{\epsilon}_{\bm{D}_{k+1}\bm{x}_{k|k}} \\ +\bm{\delta_{K_{k+1}}}\bm{{y}}_{k+1}+\bm{K}_{k+1}\bm{\epsilon_{{y}_{k+1}}}+\bm{\epsilon}_{\bm{{K}}_{k+1}\bm{{y}}_{k+1}} ,    
\end{split}
\end{equation}
where the covariance matrix $\bm{\Sigma_\times}$ of $\bm{\epsilon_\times} = \bm{\epsilon}_{\bm{D}_{k+1}\bm{x}_{k|k}}+ \bm{\epsilon}_{\bm{{K}}_{k+1}\bm{{y}}_{k+1}}$ is given by
\begin{align}
\bm{\Sigma_\times}&=\cov{\bm{\epsilon}_{\bm{D}_{k+1}\bm{x}_{k|k}}}+\cov{\bm{\epsilon}_{\bm{\Q{K}}_{k+1}\bm{\Q{y}}_{k+1}}}\\
&=\eye_c (c+d) \frac{2^{-2m}}{12},    
\end{align}
and $\cov{\bm{\epsilon}_{x_{k|k}}} = \eye_c  \frac{2^{-2m}}{12}$, $\cov{\bm{\epsilon}_{y_{k+1}}} = \eye_d  \frac{2^{-2m}}{12}$.    
Equation~\eqref{eq:delta_quanti} gives us the quantization error on the computation of $\bm{\hat{x}_{k+1|k+1}}$ based on the unquantized values of $\bm{\hat{x}_{k|k}}$, the filter parameters, and the quantization resolution $m$.

\subsection{Unreliable memory error}

We now consider the second source of noise from the unreliable memories, and derive an expression for the covariance matrix $\bm{\Gamma} = \cov{\bm{\delta \hat{x}_{k+1|k+1}^\mem}}$ of the unreliable memory noise $\bm{\delta \hat{x}_{k+1|k+1}^\mem}$ introduced in~\eqref{eq:propag_x}.

Assuming $\mathbb{E}[\gamma] \ll \operatorname{Var}[\gamma]$, as discussed in Section~\ref{sec:unreliable_Kalman}, the variance $\operatorname{Var}[\gamma] = \sigma_\gamma^2$ of the memory noise $\gamma$
can be approximated by the MSE as  $\sigma_\gamma^2  \approx \mathbb{E}[(\tilde{x}-\hat{x})^2]$. 
The value of $\mathbb{E}[(\tilde{x}-\hat{x})^2]$ depends on the error probabilities $p_b$, but also on the probability distributions of the variables $x$ which are stored in memory. 
However, from~\cite[Claim 17]{kim_generalized_2018}, if $p_{n-1} \ll \frac{1}{2}$ or $\operatorname{Pr}\left(\hat{x}_{b}=\hat{x}_{b^{\prime}}\right) \simeq \operatorname{Pr}\left(\hat{x}_{b} \neq \hat{x}_{b^{\prime}}\right)$ for
any $b \neq b^{\prime}$, then the MSE $\mathbb{E}[(\tilde{x}-\hat{x})^2]$ can be approximated as:
\begin{equation}\label{eq:MSEx}
 \sigma_\gamma^2 = \mathbb{E}[(\tilde{x}-\hat{x})^2] \approx \sum_{b=-m}^{n-1} 4^{b} p_{b} = \sum_{b=-m}^{n-1} 4^{b} e^{-e_b a} \,,
\end{equation}
where the last equality is obtained from the noise-vs-energy model~\eqref{eq:link_proba_energy}.
Therefore, the probability distributions of the variables $x$ have actually no significant impact on the value of the MSE.

Equation~\eqref{eq:MSEx} gives us a relation between the noise variance $\sigma_\gamma^2$ and the vector $\bm{e}$ of energy levels defined in~\eqref{eq:energy_vector}.
Moreover, by using~\eqref{eq:MSEx}, we show that the covariance $\bm{\Gamma}$ of the memory noise vector $\bm{\delta \hat{x}_{k+1|k+1}^\mem}$ is given by:
\begin{equation}
\bm{\Gamma} = \bm{I}_c \sigma_\gamma^2    .
\end{equation}

\subsection{Total error}
\label{sec:tot_error}

After separately studying the two error terms $\bm{\delta x_{k+1|k+1}^\quant}$ and $\bm{\delta \hat{x}_{k+1|k+1}^\mem}$, we now combine them to get an expression of the total estimation error $\bm{e^*_{k+1|k+1}} = \bm{\tilde{x}}_{k+1|k+1}-\bm{x}_{k+1}$ .
We then provide the covariance matrix $\bm{P^*}_{k+1|k+1}$ of this total error.

By using $\bm{\tilde{x}}_{k|k}$ to denote the faulty estimate of $\bm{x}_k$, we can express
\begin{equation}
    \label{eq:x_tilde_x_hat}
 \bm{\tilde{x}}_{k|k} = \bm{\hat{x}}_{k|k}+\bm{\Delta \tilde{x}_{k|k}} \,.
\end{equation} 

Considering that only the $\bm{\hat{x}}_{k|k}$ are stored in the unreliable memories, the error propagation model~\eqref{eq:propag_x} can be rewritten as:
\begin{align}
    \bm{\Delta \tilde{x}_{k+1|k+1}}  = ~ & \bm{D}_{k+1}\bm{\Delta \tilde{x}_{k|k}}+\bm{\delta \hat{x}_{k+1|k+1}^\quant}+\bm{\delta \hat{x}_{k+1|k+1}^\mem}\\ \notag
     = ~ & \bm{D}_{k+1}\bm{\Delta \tilde{x}_{k|k}}+\bm{\delta_{D_{k+1}}}\bm{\tilde{x}}_{k|k}+\bm{D}_{k+1}\bm{\epsilon}_{x_{k|k}} \\ \notag &+\bm{\delta_{K_{k+1}}}\bm{{y}}_{k+1}+\bm{K}_{k+1}\bm{\epsilon_{y_{k+1}}} \\ 
     &+\bm{\epsilon_\times}+\bm{\delta \hat{x}_{k+1|k+1}^\mem}\label{eq:deltax_1} \\ \notag
     = ~ &  (\bm{D}_{k+1}+\bm{\delta_{D_{k+1}}})\bm{\Delta \tilde{x}_{k|k}}+\bm{\delta_{D_{k+1}}}\bm{\hat{x}}_{k|k}\\ \notag
     &+\bm{D}_{k+1}\bm{\epsilon}_{x_{k|k}} + \bm{\delta_{K_{k+1}}}\bm{{y}}_{k+1}+\bm{K}_{k+1}\bm{\epsilon_{y_{k+1}}} \\ 
     &+\bm{\epsilon_\times}+\bm{\delta \hat{x}_{k+1|k+1}^\mem},  \label{eq:deltax_2}
\end{align}
where~\eqref{eq:deltax_1} is obtained by replacing $\bm{\delta \hat{x}_{k+1|k+1}^\quant}$ by its expression~\eqref{eq:delta_quanti}, and~\eqref{eq:deltax_2} comes from~\eqref{eq:x_tilde_x_hat}, which allows to write $\bm{\tilde{x}}_{k|k} = \bm{\hat{x}}_{k|k}+\bm{\Delta \tilde{x}_{k|k}}$.
Equation~\eqref{eq:deltax_2} provides a recursive form of the error at step $k$, since $\bm{\Delta \tilde{x}_{k+1|k+1}}$ depends on $ \bm{\Delta \tilde{x}_{k|k}}$ and $\bm{\hat{x}}_{k|k}$. All the other terms in~\eqref{eq:deltax_2} come from the current iteration $k+1$.

In certain conditions, such as if $\bm{H}$ and $\bm{F}$ are only composed of integer components, the total estimation error 
$
        \bm{\tilde{x}}_{k+1|k+1}-\bm{x}_{k+1} = \bm{\hat{x}}_{k+1|k+1}+\bm{\Delta \tilde{x}_{k+1|k+1}}-\bm{x}_{k+1} 
$
can be further developed as (see Appendix~\ref{sec:appendix_error} for more details):
\begin{align}
    \bm{\tilde{x}}_{k+1|k+1}-\bm{x}_{k+1} & =   (\bm{D}_{k+1}+\bm{\delta_{D_{k+1}}})(\bm{\tilde{x}}_{k|k}-\bm{x}_{k}) \\
   & +(\bm{K}_{k+1}+\bm{\delta_{K_{k+1}}})\bm{v}_{k+1} +((\bm{K}_{k+1}+\bm{\delta_{K_{k+1}}})\bm{H}-\eye)\bm{u}_k  \\ 
   & +\bm{D}_{k+1}\bm{\epsilon}_{x_{k|k}} +\bm{K}_{k+1}\bm{\epsilon_{y_{k+1}}}+\bm{\epsilon_\times} +\bm{\delta \hat{x}_{k+1|k+1}^\mem}.
\end{align}
    
This equation gives us a recursive form of the total estimation error $  \bm{\tilde{x}}_{k+1|k+1}-\bm{x}_{k+1}$ at step $k+1$,  depending on the estimation error $  \bm{\tilde{x}}_{k|k}-\bm{x}_{k}$ at step $k$,  and on the quantization resolution $2^{-m}$ and the memory noise $\bm{\delta \hat{x}_{k+1|k+1}^\mem}$.

Finally, we can compute the covariance matrix $\bm{P^*}_{k+1|k+1}$ of this error as
\begin{equation}
\label{eq:new_cov}
\begin{aligned} 
    \bm{P^*}&_{\hspace{-0.2cm}k+1|k+1}  =  \cov{\bm{\tilde{x}}_{k+1|k+1}-\bm{x}_{k+1}} \\
    = & (\bm{D}_{k+1}+\bm{\delta_{D_{k+1}}})\bm{P^*}_{k|k}(\bm{D}_{k+1}+\bm{\delta_{D_{k+1}}})^\top  \\ 
    +& (\bm{K}_{k+1}+\bm{\delta_{K_{k+1}}})\bm{R}(\bm{K}_{k+1}+\bm{\delta_{K_{k+1}}})^\top   \\  
    + &((\bm{K}_{k+1}+\bm{\delta_{K_{k+1}}})\bm{H}-\eye)\bm{Q}((\bm{K}_{k+1}+\bm{\delta_{K_{k+1}}})\bm{H}-\eye)^\top \\ 
     +& \bm{D}_{k+1}\cov{\bm{\epsilon}_{x_{k|k}}}\bm{D}_{k+1}^\top + \bm{K}_{k+1}\cov{\bm{\epsilon_{y_{k+1}}}}\bm{K}_{k+1}^\top+\bm{\Sigma_\times}+\bm{\Gamma}   ,
\end{aligned}    
\end{equation}
where all the terms involved, including the covariance matrices, have been explicited in the previous sections.  Equation~\eqref{eq:new_cov} shows that the covariance matrix $\bm{P^*}_{k+1|k+1}$  can be computed recursively.    

Equation~\eqref{eq:new_cov} provides us a measure of the performance of the filter, depending on the quantization resolution and on the energy supplied to the memory. Equipped with this derivation, we can now use the covariance matrix $\bm{P^*}_{k+1|k+1}$ as a performance criterion against which to optimize the energy consumed of the unreliable memory. 

\section{Energy optimization}\label{sec:opti}
In this section, we optimize the energy consumption of the memory while satisfying a performance constraint defined on the total estimation error of the filter.
As parameters to optimize, we consider the number of bits $m$ for the quantization,  and the  energy vector $\bm{e}$ of the memory banks. 
We define two optimization problems, which both seek to minimize the energy consumed by the memory. In the first problem, we find the optimal number of bits $m$ and the corresponding $n+m$ levels of energy to allocate to the memory banks.
Although solving this problem provides the minimum energy that needs to be supplied to the memory, it is not very practical since each of the $n+m$ bits should be stored in a different memory bank with a specific voltage supply.  
Therefore, in the second problem, we consider that the number of bits $m$ is fixed, but that the number of possible energy levels is limited to $L$ possibilities. Both the $L$ energy values and the allocation of each bit to one of the $L$ possible values should be optimized. Solving this problem allows to consider only $L<n+m$ different memory banks.

\subsection{Optimization across all the bits}
\label{sec:opti_allbits}
We first find the optimal level of energy $e_b$ of each memory bank and the optimal number of fractional bits $m$ so as to minimize the total memory energy consumption.
As performance criterion, we consider the covariance matrix  $\bm{P}^*_{N|N}$ of the total estimation error at step $N$, where $N$ is chosen to be large enough so that the filter can converge. We further introduce a matrix $\bm{\mathcal{V}}$ of the same size as $\bm{P}^*_{N|N}$ so as to define the performance constraint for the variances and covariances of estimation error on each component.
The optimization problem is then defined as follows:
\begin{equation}
    \tag{Problem 1}
\begin{aligned}
    \label{eq:problem_eqs}
    \min_{\mathbf{e},m } & \quad e_\text{tot} = \sum_{b=-m}^{n-1} e_b = \mathbbm{1}^\top \bm{e} \,,\\
\text{s.t.} &\quad \bm{P}^*_{N|N} \prec \bm{\mathcal{V}}  \; \text{and} \; e_b \geq e_{\text{thres}} \  \forall b \in  \llbracket-m,n-1 \rrbracket \,,
\end{aligned}
 \end{equation}
where $\prec$ is a component-wise inequality between the two matrices, and where the minimum is taken over all energy vectors $\bm{e}$ as defined in (\ref{eq:energy_vector}) and for all the possible values of the number of bits  $m$.
We consider that $m\in \llbracket 0,M\rrbracket$, where $M$ is the maximum number of bits which could be stored in a memory.
The value $e_{\text{thres}}$ is the minimum level of energy for each memory bank, so as to avoid undesired effects such as circuit delays and energy leakage~\cite{dreslinski_near-threshold_2010}.

\ref{eq:problem_eqs} involves one discrete parameter $m$ and $m+n$ continous parameters $\bm{e}$ which makes it hard to solve at once.
As a first step, we assume that the value of $m$ is fixed, and solve the following simplified problem:
\begin{equation}
 \tag{Problem 1-a}
 \begin{aligned}
    \label{eq:problem_eq_mfixed}
    \min_{\mathbf{e}}  & \quad e_\text{tot} = \sum_{b=-m}^{n-1} e_b = \mathbbm{1}^\top \bm{e} \,, \quad \\
\text{s.t.} \; & \bm{P}^*_{N|N} \prec \bm{\mathcal{V}}  \quad \text{and} \quad e_b \geq e_{\text{thres}} \  \forall b \in  \llbracket-m,n-1 \rrbracket \,,
\end{aligned}
\end{equation}
 by using the Karush–Kuhn–Tucker (KKT) conditions (see Appendix~\ref{sec:appendix_KKT}). 
From these conditions, we show that the optimal energy level $e_b^*$ for bit $b$ has expression:
\begin{equation}
\label{eq:opt_solution}
  e_b^*=\begin{cases}
    e_{\text{thres}}, & \text{if $\lambda<\frac{1}{4^ba}$} \,,\\
    \frac{1}{a}\log(4^b a \lambda), & \text{otherwise} \,,
  \end{cases}
\end{equation}
where $\lambda$ is a dual variable that balances the trade-off between reducing the energy consumption and preserving the performance of the system.
The optimal vector $\bm{e}^*$ can then be computed using a water-filling algorithm \cite{kim_generalized_2018} for a fixed desired performance $\bm{\mathcal{V}}$ of the filter. With this optimal solution, we see that the least significant bits have their energy adjusted to the minimum possible energy level $e_\text{thres}$. The energy levels then increase logarithmically for each bit as their significance increase.

\begin{algorithm}[t!]
\label{Alg:opti_all}
\SetAlgoLined
\hspace*{0.05cm} \textbf{Input:} $\bm{\mathcal{V}}$, $a$, $\beta$, $\xi$, $e_\text{thres}$ \;
  $\bm{e}_\text{min}$ $\xleftarrow{}$ $+\infty$\;
\For{each value of m}{
   $\bm{e}$ $\xleftarrow{}$ $e_{\text{thres}}$ \;
   $\bm{P}_\text{prev} = 0 $\;
 \While{$\bm{P}^*_{N|N}(\bm{e},m) \succ \bm{\mathcal{V}}$ and $\left\| \bm{P}_\text{prev}-\bm{P}^*_{N|N}  \right\| > \xi$}{
     $\bm{P}_\text{prev} = \bm{P}^*_{N|N}(\bm{e},m)$\;
  $b$ $\xleftarrow{}$ $\underset{b}{\arg\,\min}$  $\{\log(\frac{1}{4^ba})+e a\}$\;
  $e_b$ $\xleftarrow{}$ $e_b+\beta$  \;
  }
  \If{$\bm{P}^*_{N|N}(\bm{e},m) \prec \bm{\mathcal{V}}$ and $\sum_{b=-m_\text{opti}}^{n} e_{\text{min}_b} > \sum_{b=-m}^{n} e_b$}{
  $\bm{e}_\text{min}$ $\xleftarrow{}$ $\bm{e}$ \;
  $m_\text{opti}$ $\xleftarrow{}$ $m$ \;
  }
 }
 \KwResult{Optimal number of bits $m_\text{opti}$ and optimal energy allocation vector $\bm{e}_\text{min}$}
 \caption{Computing the optimal values for $\bm{e}$ and $m$}
\end{algorithm}

Since $m$ is discrete, the optimal solution~\eqref{eq:opt_solution} is computed using the water-filling algorithm for each possible value of $m$. We then retain the solution $(m^\star,\bm{e}^{\star})$ which gives the lowest total energy $e_\text{tot}^\star = \sum_{b=-m^\star}^{n} e^\star_b$.
In this method, the influence of the quantization error is taken into account through the performance criterion $\bm{P}^*_{N|N}$.
For a small number of bits $m$, quantization errors may make it impossible to satisfy the desired performance constraint and therefore the water-filling algorithm will not be able to find an optimal solution.
In this case, if we detect that the algorithm converges toward a performance value that is still higher than the constraint, the algorithm is stopped and we proceed to the next value of $m$ in the considered range.

The full optimization process is summarized in Algorithm~\ref{Alg:opti_all}.
In this algorithm, the parameter $\beta$ controls the rate at which the energy for each memory bank is increased at each iteration. The value of $\beta$ is chosen either using the precision with which energy can be set in a given device technology, or based on the desired rate of convergence for the water-filling algorithm.
The condition $\left\| \bm{P}_\text{prev}-\bm{P}^*_{N|N}  \right\| > \xi$ is used to detect whether the water-filling algorithm has a feasible solution, so the value of $\xi$ is set to be low.

\subsection{Optimization with a limited number of energy levels}
\label{sec:opti_levels}
%
In practice, the solution of Problem $1$ makes the implementation costly as each 
bit position should be stored in a separate memory bank.
Therefore, we define a second optimization problem with only $L<m+n$ possible levels of energy. For implementation purposes,  we only consider small values for $L$ ($L<10$).  
The vector $\bm{\newenerg} = [\newenerg_0, \hdots, \newenerg_{L-1}]$ contains the $L$ levels of energy. We use $n_\ell$ to denote the number of bits allocated to energy level $\newenerg_l$, so that $\sum_{\ell=0}^{L-1} n_\ell = n+m$. This means that each memory bank of the energy group $\ell$ has an energy level $e_b = \frac{\newenerg_\ell}{n_\ell}$. We write $\bm{n} = [n_0,\hdots,n_{L-1}]$ for the vector containing the $L$ values $n_\ell$.

In the following, for simplicity, we consider that the number of bits $n$ and $m$ are fixed, and we seek to optimize the total energy consumption of the unreliable memory for a fixed number of energy levels $L$. The objective is to reduce the total energy consumed by the unreliable memory by allocating different levels of energy to the $L$ groups of bits. Two parameters are considered in this optimization: the values of each energy level $e_\ell$ and the number of bits allocated to each of these energy levels $n_\ell$.
The optimization problem can be written as:
\begin{equation}
    \tag{Problem 2}
\begin{aligned}
    \label{eq:problem_eqs_levels_general}
    \min_{\bm{\newenerg},\bm{n}} & \quad e_\text{tot} = \sum_{l=0}^{L-1} \newenerg_\ell = \mathbbm{1}^\top \bm{\newenerg} \,,\\
\text{s.t.} &\quad \bm{P}^*_{N|N} \prec \bm{\mathcal{V}}  \quad \text{and} &\newenerg_\ell \geq e_{\text{thres}} n_\ell  ,  \; n_\ell \in  \llbracket1,n+m-L \rrbracket \, \forall \ell \in  \llbracket0,L-1 \rrbracket \,,
\end{aligned}
 \end{equation}
First, we solve the optimization problem in the case where we know which bit is allocated to which energy level. This means that the values of $n_\ell$ are known and that we only want to compute the optimal values of the energy levels $\newenerg_\ell$. In this case the optimization problem can be written as:
\begin{equation}
    \tag{Problem 2-a}
\begin{aligned}
    \label{eq:problem_eqs_levels}
    \min_{\bm{\newenerg}} & \quad e_\text{tot} = \sum_{l=0}^{L-1} \newenerg_\ell = \mathbbm{1}^\top \bm{\newenerg} \,,\\
\text{s.t.} &\quad \bm{P}^*_{N|N} \prec \bm{\mathcal{V}}  \quad \text{and} \quad \newenerg_\ell \geq e_{\text{thres}} n_\ell \quad  \forall \ell \in  \llbracket0,L-1 \rrbracket \,,
\end{aligned}
 \end{equation}
        
\ref{eq:problem_eqs_levels} is quite similar to the one described in Section~\ref{sec:opti_allbits} and can be solved using the same method as the one presented in Appendix~\ref{sec:appendix_KKT}, by relying on the KKT conditions. The optimal solution in this case is:
\begin{equation}
\label{eq:opt_solution_levels}
  \newenerg_l^*=\begin{cases}
        e_{\text{thres}} n_\ell, & \text{if $\lambda<\frac{1}{\sum_{b=0}^{n_\ell} 4^ba}$} \,,\\
    \frac{1}{a}\log(\sum_{b=0}^{n_\ell} 4^b a \lambda), & \text{otherwise} \,,
  \end{cases}
\end{equation}

This solution allows us to compute the optimal energy levels $\newenerg_\ell$ for a given energy allocation across the bits. The second step consists of computing the best allocation of bits to each energy group. Given that we only consider small values of $L$, we compute the optimal solution from~\eqref{eq:opt_solution_levels} for each possible energy allocation of the bits. Then the solution with the smallest total energy $\sum_{\ell=0}^{L-1} \newenerg_{\ell} $ is retained.
Although Problem $2$ leads to more practical solution, it is expected that the optimal total energy of the memory is higher for Problem $2$ than for Problem $1$.  


\section{Simulation results}\label{sec:simu}
 In our simulations, unless explicitly stated, we consider a simple tracking problem where the state vector $\bm{x}$ is composed of two variables representing the position and velocity of an object. Measurements $y$ only consist of noisy observations of the position of the object. 
The process matrix $\bm{F}$ and measurement matrix $\bm{H}$ are defined as
\begin{equation}
    \bm{F} = \begin{bmatrix} 1 & \dt \\ 
                              0 & 1\end{bmatrix} \,, \,
    \bm{H} = \begin{bmatrix} 1 & 0 \end{bmatrix} \,,
\end{equation}
and the process noise covariance matrix $\bm{Q}$ and measurement covariance matrix $\bm{R}$ are given by
\begin{equation}
    \bm{Q} = \begin{bmatrix} \sigma_x^2 & 0 \\ 
                              0 & \sigma_x^2\end{bmatrix} \,,\,
    \bm{R} = \begin{bmatrix} \sigma_y^2 \end{bmatrix} \,.
\end{equation}
where $\dt=1$ and $\sigma_x = 0.01$ and $\sigma_y = 10$. The factor $a$ in~\eqref{eq:link_proba_energy} is taken as $a=12.8$ as in~\cite{hacene_training_2019}.
This section is divided into two parts. We first evaluate the accuracy of the proposed theoretical analysis, and we then provide solutions to the two considered optimization problems.

\subsection{Accuracy of the theoretical analysis}
\begin{figure}[t]
  \centering
\includegraphics[width=0.8\linewidth]{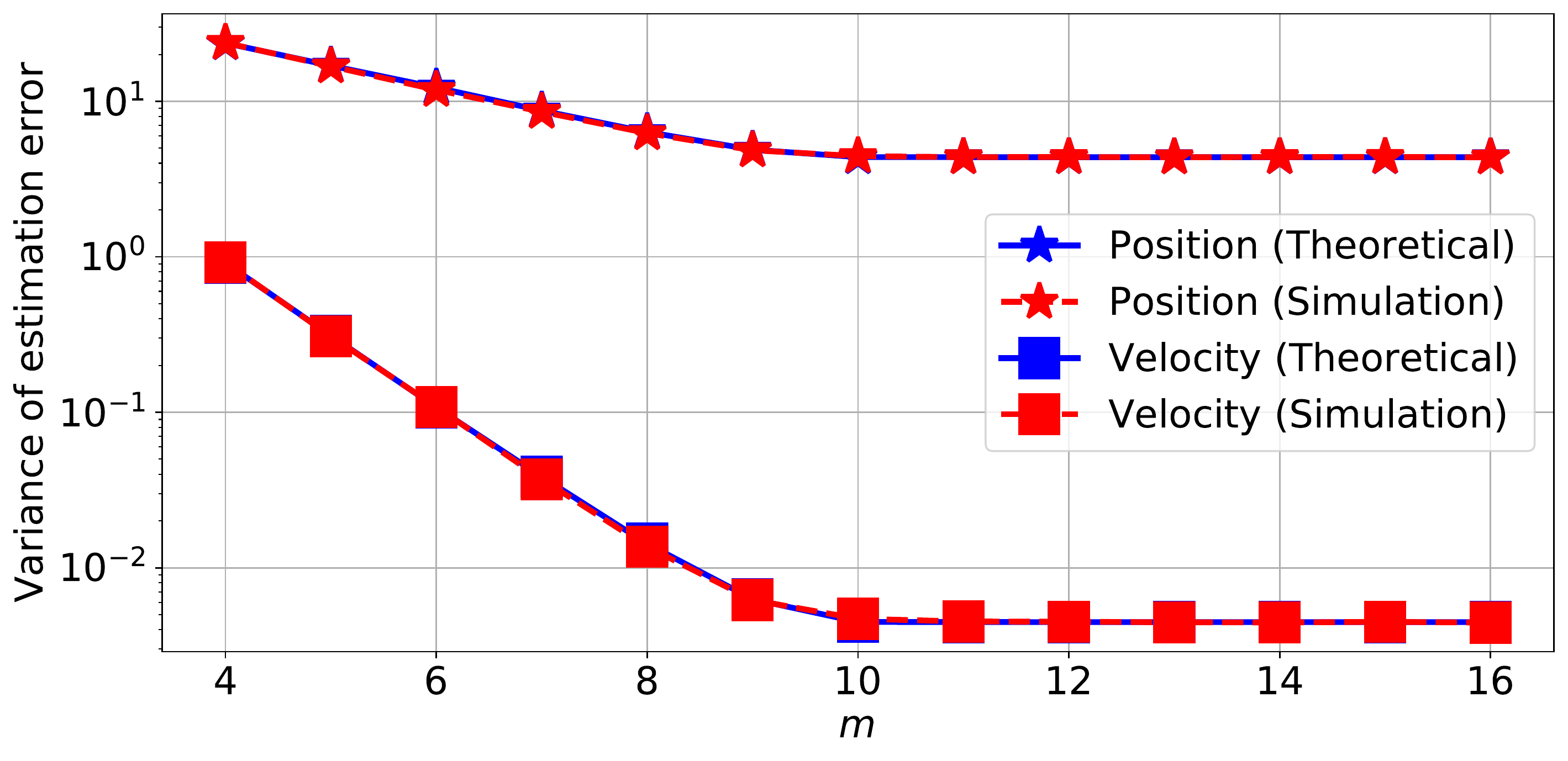}
\caption{Theoretical and simulated variance of estimation error on the position and velocity depending on the number of quantization bits, using a reliable memory.}
\label{fig:quanti}
\end{figure}

First, to evaluate the accuracy of the proposed theoretical analysis, we perform Monte Carlo simulations ($N_{mc}=10^7$) and measure the covariance matrix of the error on the estimation at step $N=250$, thus giving enough time for the filter to converge in normal conditions. This covariance matrix is compared with the theoretical expression of the covariance of the estimation error $\bm{P^*}_{N|N}$ computed in Section~\ref{sec:tot_error}.
Figure~\ref{fig:quanti} shows the variance of the estimation errors on the position and the velocity for different values of $m$ in the case of a reliable memory, meaning that we consider only the quantization error and not the memory noise. We observe that the theoretical predictions of the errors closely match the simulations, which shows the accuracy of our theoretical analysis. Moreover, from Figure~\ref{fig:quanti} we can observe that for a small number of bits $m$ the quantization error is large and dominates the total estimation error. However, starting from $m=10$ bits the estimation error reaches a constant level which can be interpreted as a minimum bound on the estimation error that one can obtain from a standard Kalman filter for this tracking problem. This shows that from $m=10$ bits, the quantization errors become negligible compared to the estimation error achieved by the standard full precision Kalman filter. 
Thus at this point using more bits will not result in minimizing the estimation error, justifying the need for optimizing the parameter $m$.

\begin{figure}[t]
  \centering
\includegraphics[width=1.0\linewidth]{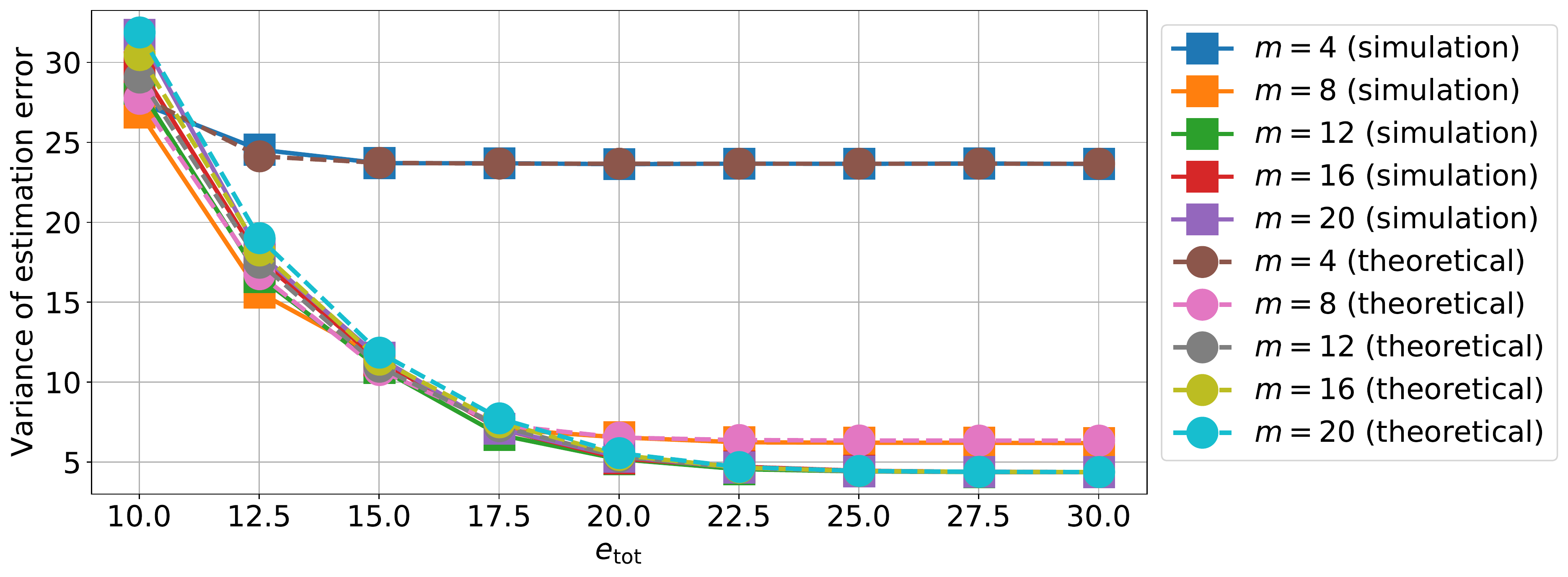}
\caption{Theoretical and simulated variance of estimation error on the position depending on the supplied energy to each number using an unreliable memory for different value of number of quantization bits $m$.}
\label{fig:var_bits_a}
\end{figure}

\begin{figure}[t]
\centering
\includegraphics[width=0.8\linewidth]{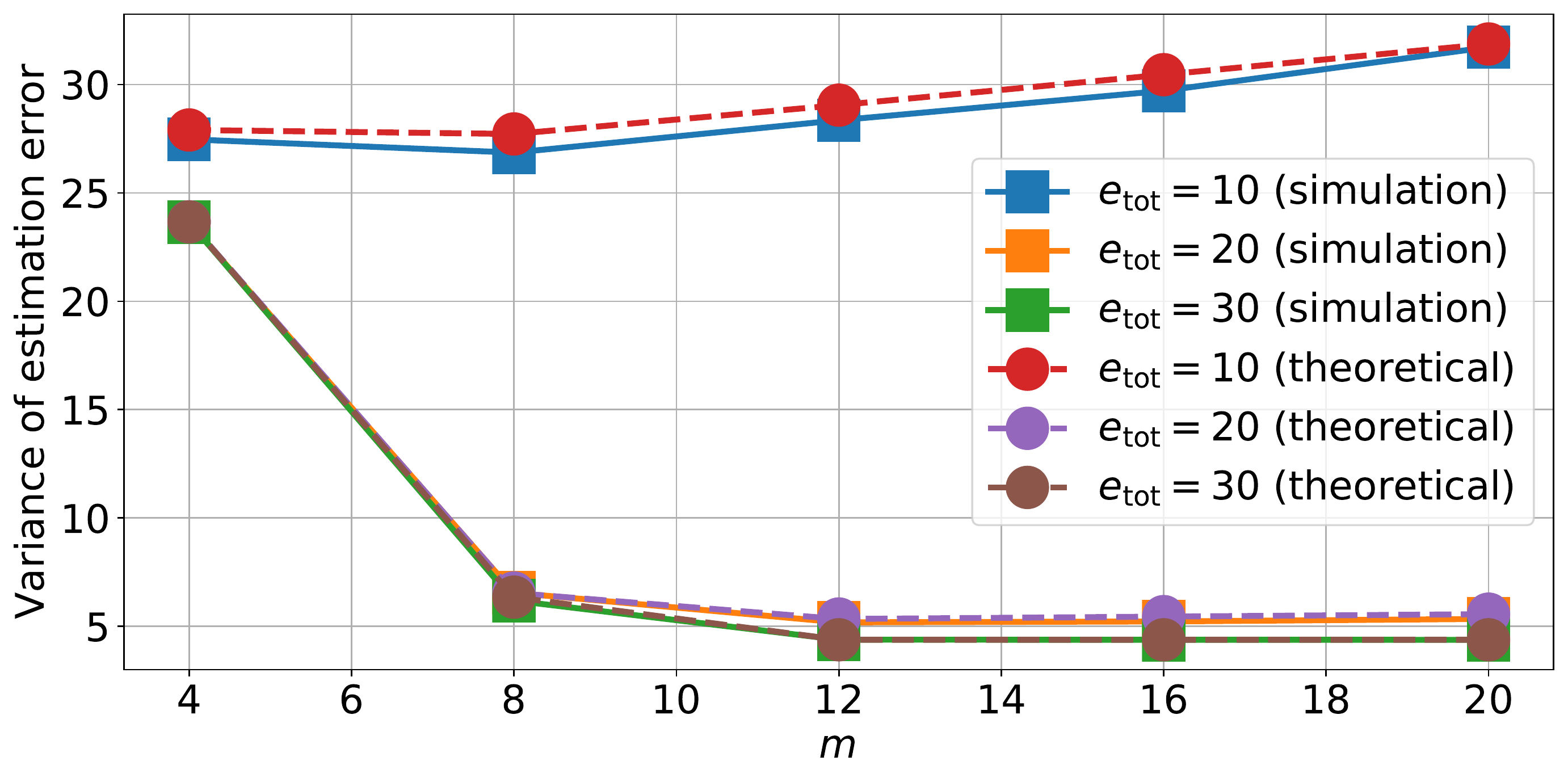}
\caption{Theoretical and simulated variance of estimation error on the position depending on the number of quantization bits using an unreliable memory for different value of energy per number $e$.}
\label{fig:var_bits_b}
\end{figure}

In a second step, we introduce the memory noise in addition to the quantization.
Figure~\ref{fig:var_bits_a} shows the variance of the estimation error on the position depending on the total energy $e_\text{tot}$ for different values of $m$. The variance values were obtained from both Monte Carlo simulations and from the theoretical analysis of Section~\ref{sec:tot_error}.  Further, Figure~\ref{fig:var_bits_b} shows the variance of estimation error on the position depending on the number of bits $m$ for different values of the total energy $e_\text{tot}$. The comparison between theoretical results and Monte Carlo simulations show the accuracy of the theoretical analysis that predicts the new computed covariance $\bm{P^*}_{k|k}$.

From Figure~\ref{fig:var_bits_a}, we can also see that both the number of bits and the total energy can affect the variance of the estimation error. If the number of bits or the energy supplied is too low then the quantization error or the memory noise will dominate the total estimation error. However we can see that there is a minimum number of bits, around $m=12$, from which given enough energy it will be possible to reach the minimum possible variance of estimation error. 
Moreover from Figure~\ref{fig:var_bits_b}, we can see that for a low value of supplied energy per variable, the variance of the estimation error will increase with the number of bits as there is too little energy. But for a larger amount of energy $e_\text{tot}>10$, the variance of the estimation error will decrease with the number of bits since the quantization error decreases. Finally for a large enough number of bits, the total estimation error will only depend on the total energy and on the variance of the estimation error of a reliable full precision filter. 

As the work presented in this paper is done to reduce the energy consumption of the memory of a Kalman filter, it is of a greater utility when the memory is large. For this reason, the results presented before were also tested on a larger Kalman filter with a dimension for the state vector $x$ of $c=20$. On these simulations we use a state transition model that performs a shifting of the entries of the state to the next state at each iteration such as the one used in~\cite{berberidis_data_2017}. That is,
\begin{equation}
      \bm{F}_{i,j}=\begin{cases}
    1, &\text{if $i-j=1$} \,,\\
    0, & \text{otherwise} \,,
  \end{cases}
\end{equation}
and $\bm{F}_{c,1} = 1$. The initial state vector is drawn from a normal distribution.

\begin{figure}[t]
  \centering
\includegraphics[width=1.0\linewidth]{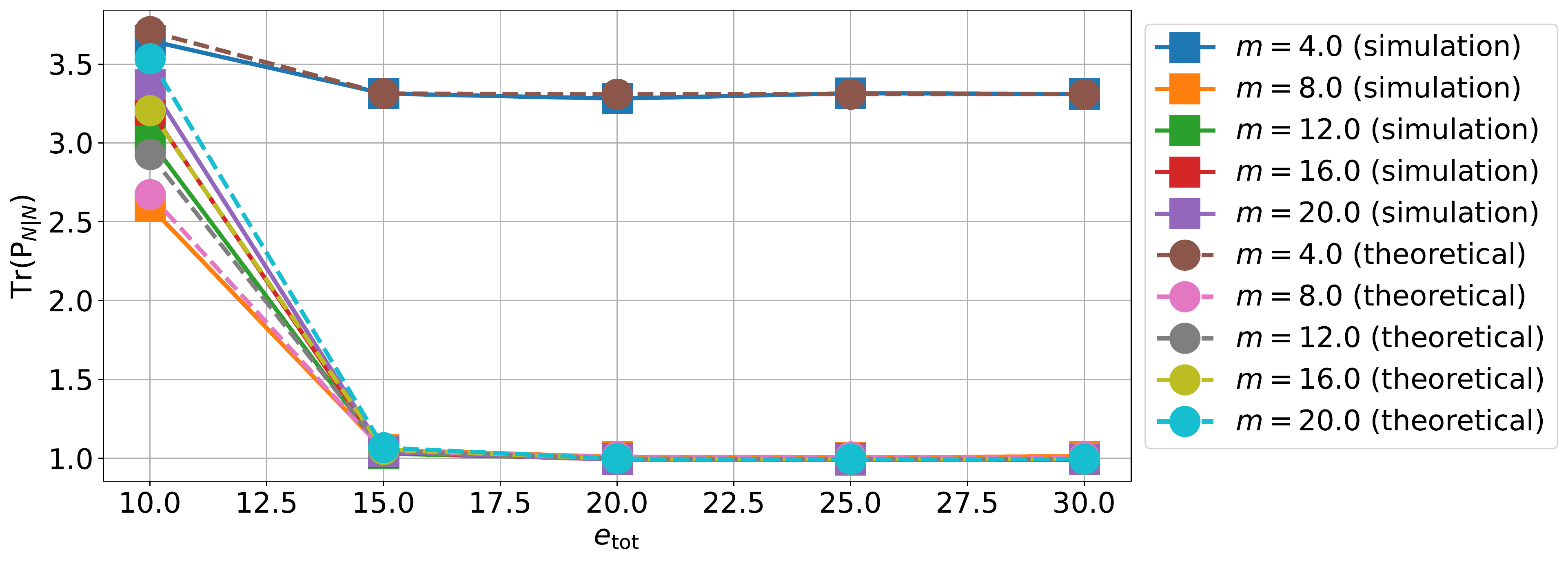}
\caption{Theoretical and simulated variance of estimation error on the position depending on the supplied energy to each number using an unreliable memory for different value of number of quantization bits $m$ in the case of a Kalman filter with a larger dimension.}
\label{fig:var_bits_a_large}
\end{figure}

\begin{figure}[t]
\centering
\includegraphics[width=0.8\linewidth]{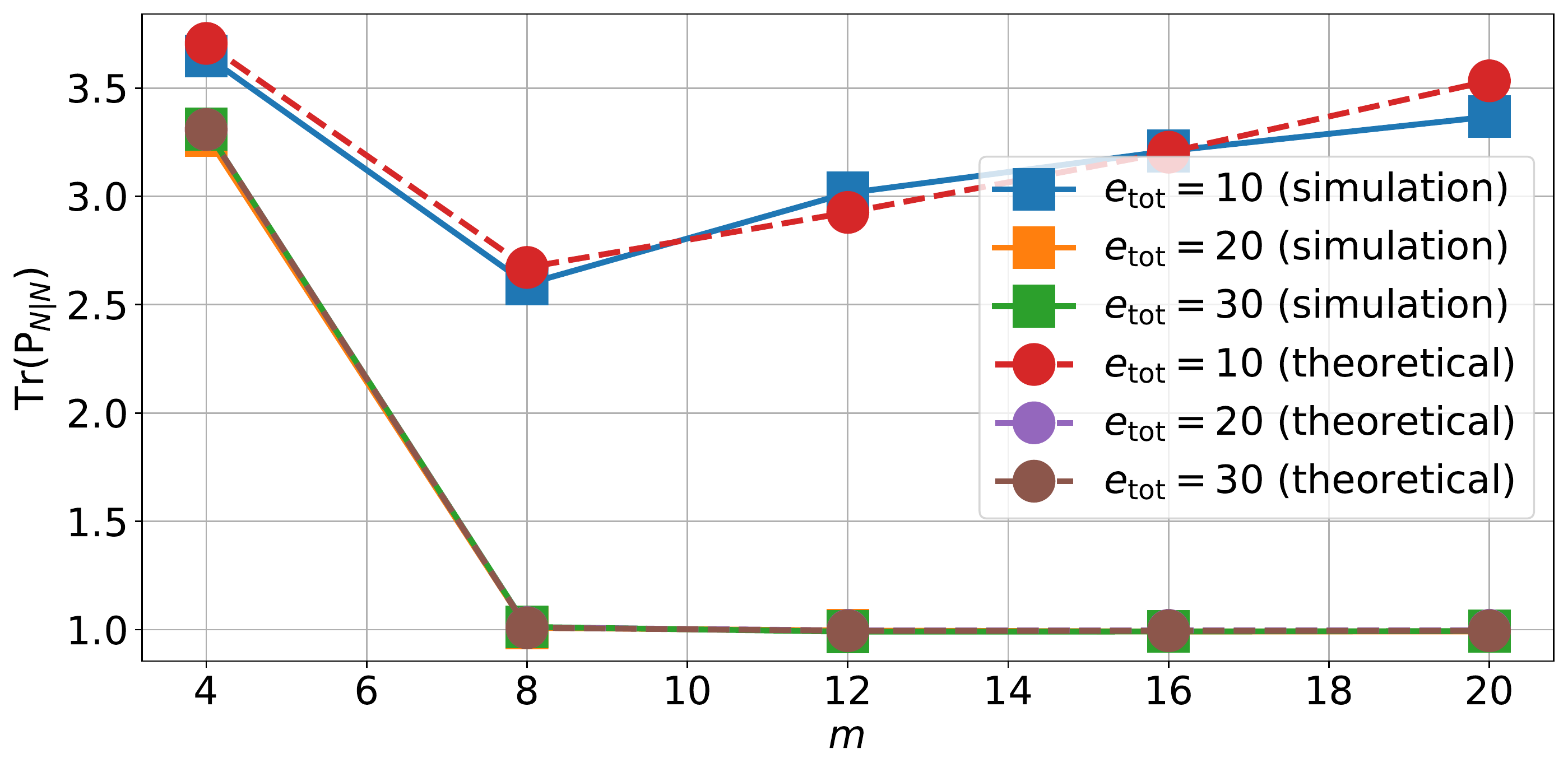}
\caption{Theoretical and simulated variance of estimation error on the position depending on the number of quantization bits using an unreliable memory for different value of energy per number $e$ in the case of a Kalman filter with a larger dimension.}
\label{fig:var_bits_b_large}
\end{figure}

In this case, the performance of the filter is measured by the trace of the covariance matrix $\bm{P}_{N|N}$. 
The results in Figures~\ref{fig:var_bits_a_large} and~\ref{fig:var_bits_b_large} show that the same conclusion can be taken from the simulations done on the smaller-scale Kalman filter and that the method presented in this paper can therefore be applied to large-size filters. 

\subsection{Solutions to the optimization problems}

\begin{figure}[t]
  \centering
\includegraphics[width=0.8\linewidth]{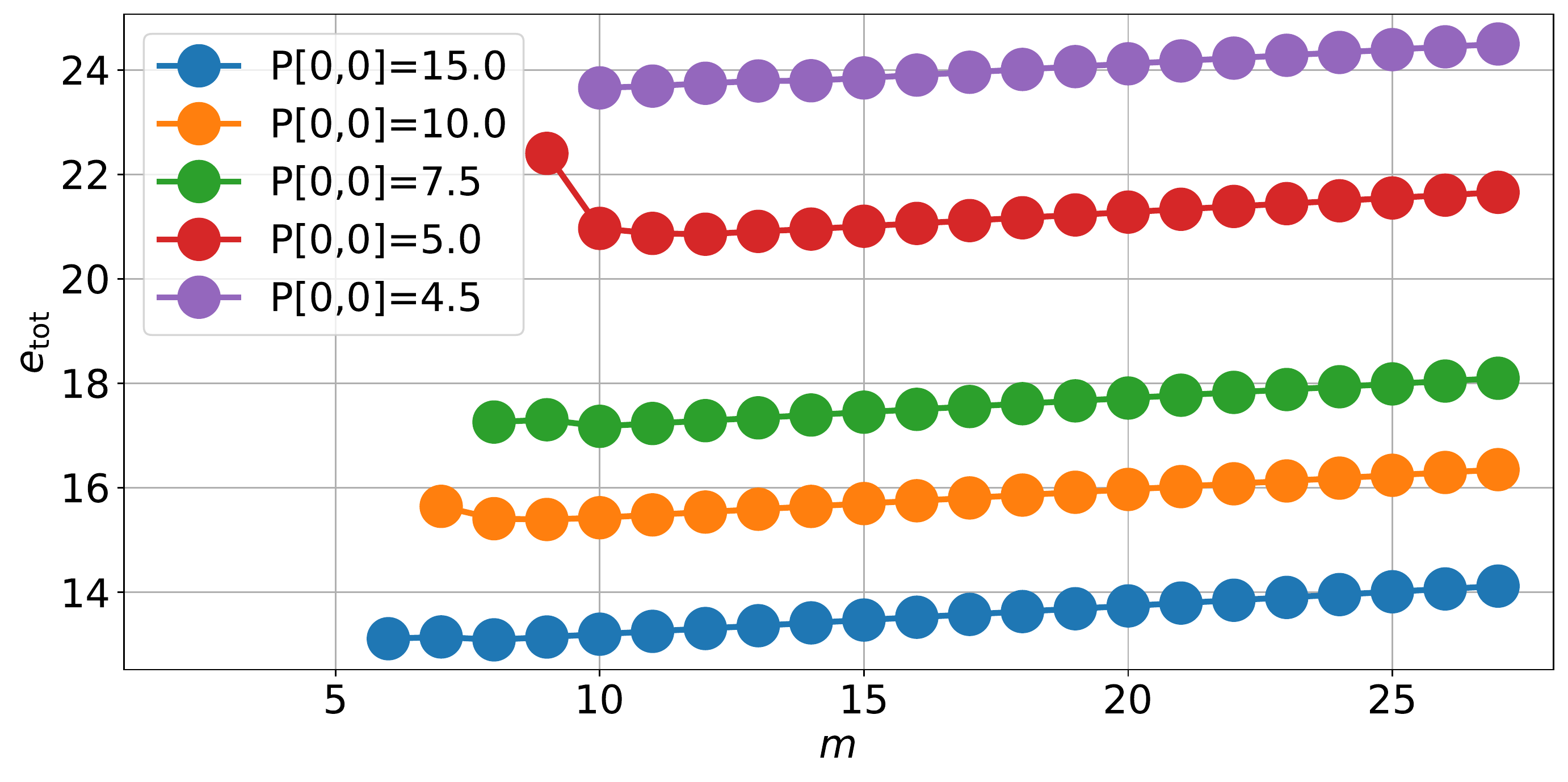}
\caption{Energy needed to store each number in an  unreliable memory to achieve various desired variance of estimation error on the position depending on the number of bits with the optimal energy allocation}
\label{fig:fixed_cov_a}
\end{figure}

\begin{figure}[t]
\begin{minipage}[h]{1\linewidth}
  \centering
  \centerline{\includegraphics[width=0.8\linewidth]{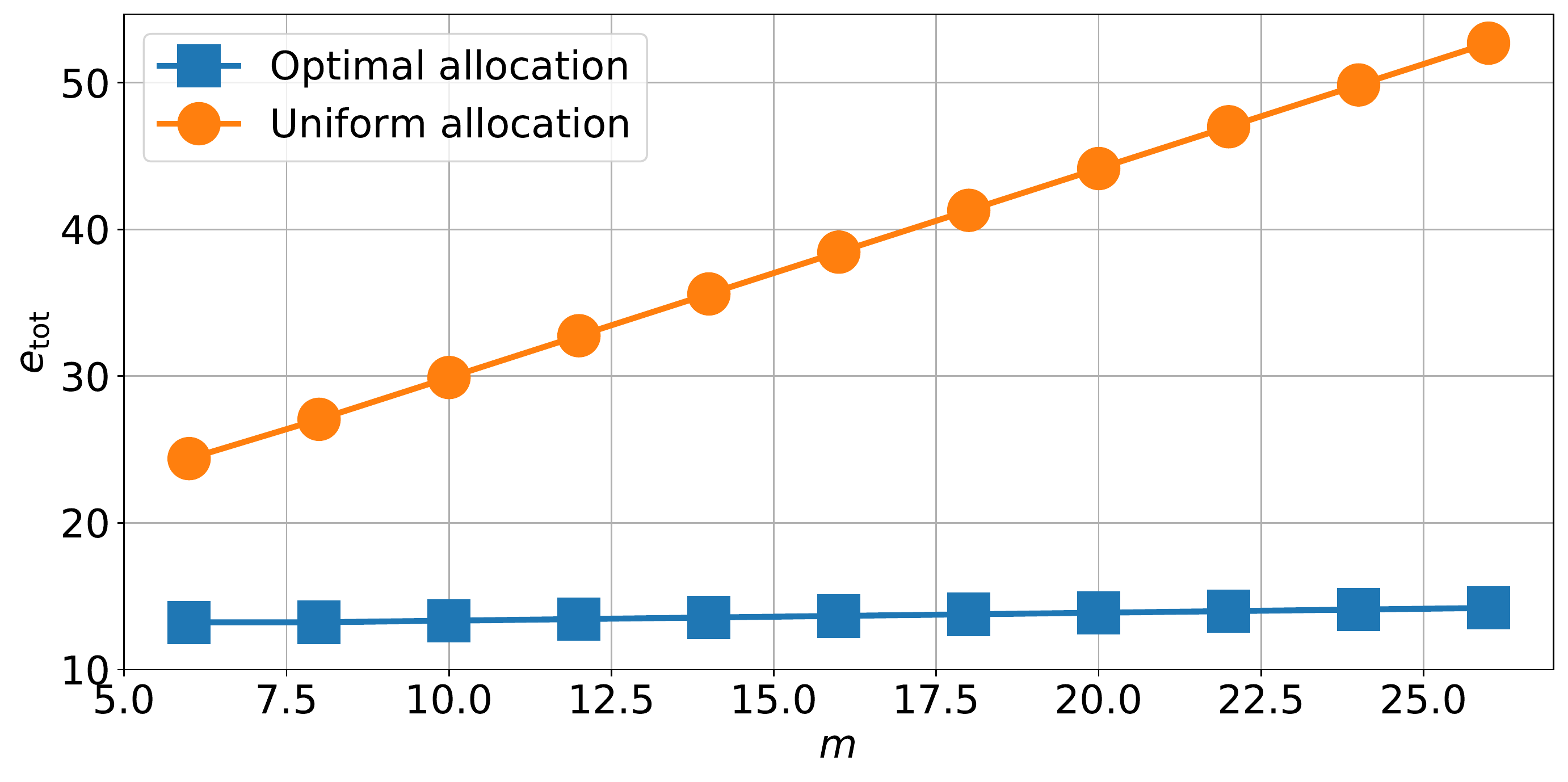}}
\caption{Energy needed to store each number in an  unreliable memory to achieve a  variance of estimation error on the position $P_{N|N}[0,0]=15$ depending on the number $m$ of bits.} 
\label{fig:fixed_cov_b}
\end{minipage}
\end{figure}

We now focus on the optimization problems introduced in Section~\ref{sec:opti}, starting with the first one.
Figure~\ref{fig:fixed_cov_a} shows the amount of energy $e_\text{tot}$ needed to store each number in the unreliable memory so as to achieve a fixed variance of estimation error on the position for each value of $m$. The total energy $e_\text{tot}$ was calculated both using the optimal allocation from Algorithm~\ref{Alg:opti_all}, and using a uniform energy allocation. From Figure~\ref{fig:fixed_cov_b}, we can see that the total energy $e_{tot}$ of the memory slightly increases with the number of bits $m+n$.  The slight increase in memory consumption is due to the form of the optimal solution~\eqref{eq:opt_solution}. Indeed, once the minimum number of bits needed to achieve the performance constraint is reached, then additional bits will be set at the minimum energy threshold $e_\text{thres}$.

Figure~\ref{fig:fixed_cov_b} compares the optimal solution from Figure~\ref{fig:fixed_cov_a} with a uniform energy allocation. This shows that the optimal energy allocation allows for a significant energy gain compared to the uniform allocation.
Here for the minimum number of bit needed to achieve the performance constraint, the optimal allocation require 56\% less energy than the uniform allocation. 

%
%
%


%
%

\begin{figure}[!t]
  \centering
 \includegraphics[width=0.8\linewidth]{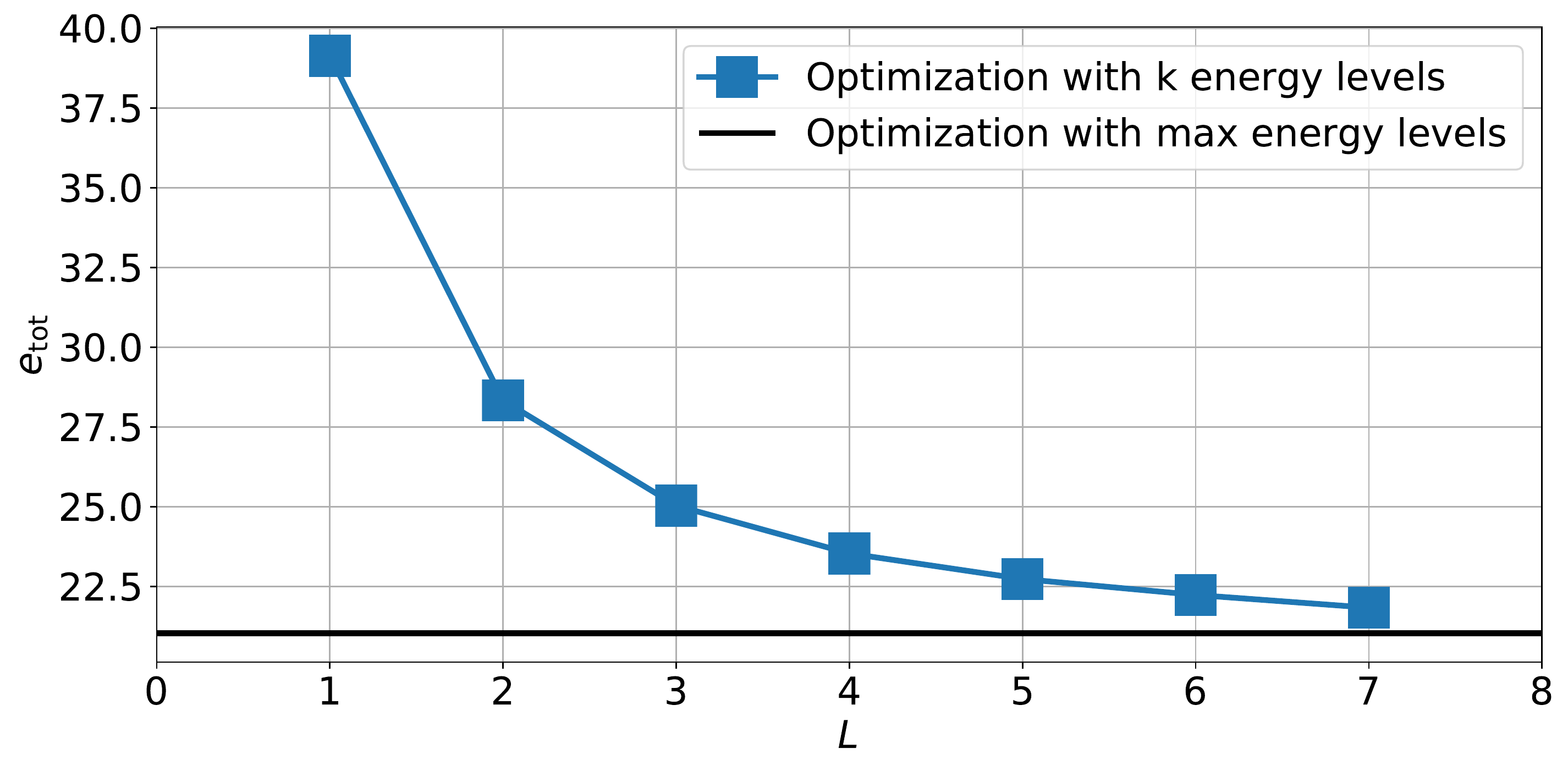}
\caption{Energy needed to store a variable in memory $e$ for different value of energy level available $L$ to achieve a fixed covariance value.}
\label{fig:fixed_levels}
\end{figure}

We now focus on the second optimization problem defined in Section~\ref{sec:opti_levels}~\eqref{eq:problem_eqs_levels} where only a limited number of energy levels are available. Figure~\ref{fig:fixed_levels} shows the total energy needed for each variable in memory to achieve a fixed level of error depending on the number $L$ of energy level possible. 
For each considered number of level $L \in \llbracket 1 , 7 \rrbracket$, the total energy $e_\text{tot}$ was computed for all possible energy allocations using the optimal solution~\eqref{eq:opt_solution_levels}. The minimum energy possible for each value of $L$ was then kept and is shown in Figure~\ref{fig:fixed_levels}. This minimum energy is compared with the minimum energy needed for~\ref{eq:problem_eqs} where there are as many energy levels possible as the number of bits. Here the total number of quantization bits is $B=20$.
We observe that even a small number of energy levels $L$ can lead to significant gains in energy. In this case, only 7 levels of energy allow to achieve 95\% of the maximum energy gain that was obtained in the first optimization problem.
When looking at the optimal energy allocation for each value of $L$ bit by bit, we notice that in most cases the optimal solution seems to be when the energy levels are uniformly shared between the bits. This means that if there are $B$ bits and $L$ levels available and $L$ is a divisor of $B$ then each group of bits assigned to each energy level will have a size of $n_\ell = \frac{B}{L}$.

\section{Conclusion}
In this paper we studied a quantized Kalman filter implemented with unreliable memories. We provided analytical expressions for the covariance matrix of the estimation error, and provided updated filter equations to take into account all considered sources of errors. We proposed and solved two optimization problems that allow to find the best trade-offs between energy consumption and performance of the filter. Simulation results showed the accuracy of the theoretical analysis and illustrated the significant energy gains provided by our approach. 
Due to the generic nature of the considered error propagation model, these results could be used for various realistic noise-versus-energy models of unreliable components. 
%


\appendix 
\subsection{Computation of  $\bm{\tilde{x}}_{k+1|k+1}-\bm{x}_{k+1}$}
\label{sec:appendix_error}
\begin{align}
\begin{split}
    \bm{\tilde{x}}_{k+1|k+1}-\bm{x}_{k+1} &= \bm{\hat{x}}_{k+1|k+1}+\bm{\Delta \tilde{x}_{k+1|k+1}}-\bm{x}_{k+1}\\
\end{split} \\
\begin{split}
    &= (\bm{D}_{k+1}+\bm{\delta_{D_{k+1}}})\bm{\Delta \tilde{x}_{k|k}}+(\bm{D}_{k+1}+ \bm{\delta_{D_{k+1}}})\bm{\hat{x}}_{k|k}+\bm{D}_{k+1}\bm{\epsilon}_{x_{k|k}}\\
    &+(\bm{K}_{k+1}+\bm{\delta_{K_{k+1}}})\bm{{y}}_{k+1} +\bm{K}_{k+1}\bm{\epsilon_{y_{k+1}}}+\bm{\epsilon_\times}+\bm{\delta \hat{x}_{k+1|k+1}^\mem}-\bm{x}_{k+1} \\
\end{split} \\
    \begin{split}
    &= (\bm{D}_{k+1}+\bm{\delta_{D_{k+1}}})\bm{\Delta \tilde{x}_{k|k}}+(\bm{D}_{k+1}+ \bm{\delta_{D_{k+1}}})\bm{\hat{x}}_{k|k}+\bm{D}_{k+1}\bm{\epsilon}_{x_{k|k}}\\
    &+(\bm{K}_{k+1}+\bm{\delta_{K_{k+1}}})(H\bm{x}_{k+1}+\bm{v}_{k+1}) +\bm{K}_{k+1}\bm{\epsilon_{y_{k+1}}}+\bm{\epsilon_\times}+\bm{\delta \hat{x}_{k+1|k+1}^\mem}-\bm{x}_{k+1} \\
    \end{split} \\
    \begin{split}
    &= (\bm{D}_{k+1}+\bm{\delta_{D_{k+1}}})\bm{\Delta \tilde{x}_{k|k}}+(\bm{D}_{k+1}+ \bm{\delta_{D_{k+1}}})\bm{\hat{x}}_{k|k}+\bm{D}_{k+1}\bm{\epsilon}_{x_{k|k}}\\
    &+(\bm{K}_{k+1}+\bm{\delta_{K_{k+1}}})\bm{v}_{k+1} +\bm{K}_{k+1}\bm{\epsilon_{y_{k+1}}}+\bm{\epsilon_\times}+\bm{\delta \hat{x}_{k+1|k+1}^\mem} \\
    &+ ((\bm{K}_{k+1}+\bm{\delta_{K_{k+1}}})H-\eye)\bm{x}_{k+1} \\
    \end{split} \\
    \begin{split}
    &= (\bm{D}_{k+1}+ \bm{\delta_{D_{k+1}}})\bm{\tilde{x}}_{k|k}-(\eye-(\bm{K}_{k+1}+\bm{\delta_{K_{k+1}}})\bm{H})\bm{F}\bm{x}_{k}\\
    &+(\bm{K}_{k+1}+\bm{\delta_{K_{k+1}}})\bm{v}_{k+1} +\bm{K}_{k+1}\bm{\epsilon_{y_{k+1}}}+\bm{\epsilon_\times}+\bm{\delta \hat{x}_{k+1|k+1}^\mem}\\
    &+\bm{D}_{k+1}\bm{\epsilon}_{x_{k|k}} +((\bm{K}_{k+1}+\bm{\delta_{K_{k+1}}})H-\eye)\bm{u}_k \,.
    \end{split}
\end{align}

If $\bm{H}$ and $\bm{F}$ are only composed of integer components, due to how the quantization is done:
\begin{equation}
    (\bm{D}_{k+1}+ \bm{\delta_{D_{k+1}}})=(\eye-(\bm{K}_{k+1}+\bm{\delta_{K_{k+1}}})\bm{H})\bm{F} \,.
\end{equation}

\subsection{Computation of the optimal solution to~\ref{eq:problem_eqs}}
\label{sec:appendix_KKT}
From optimization Problem $1$, we can define the Lagrangian:
\begin{equation}
    L(e,\nu,\lambda) =  \sum_{b=0}^{B-1} e_b + \nu(\sum_{b=0}^{B-1} 4^{b} e^{-e_b a}   - \mathcal{V}) - \sum_{b=0}^{B-1} \lambda_b (e_b - e_\text{thres}) \,.
\end{equation}

From the KKT conditions, for the optimal solution $\bm{e^*}$:

\begin{equation}
    \nu(\sum_{b=0}^{B-1} 4^{b} e^{-e^*_b a}   - \mathcal{V}) = 0  \ , \nu \geq 0
\end{equation}

\begin{equation}
    \label{eq:KKKT_2}
     \lambda_b (e^*_b - e_\text{thres}) = 0    \ , \lambda_b \geq 0  \  \forall b \in  [\![0,B-1 ]\!]
\end{equation}

\begin{equation}
   \label{eq:KKKT_3}
   \frac{\partial L}{\partial e^*_b} = 1 - \nu 4^b a e^{-e_b a}  -\lambda_b = 0
\end{equation}

From ~\eqref{eq:KKKT_2} and~\eqref{eq:KKKT_3}:
\begin{equation}
  \lambda_b  = 1 - \nu 4^b a e^{-e_b a}  \geq 0 \,.
\end{equation}

If $\nu = 0$ then $\lambda_b = 1$ and $e_b = e_\text{thres}$. Therefore, we claim that $\nu \neq 0$ and so $\sum_{b=0}^{B-1} 4^{b} e^{-e^*_b a}   = \mathcal{V}$.

If $\nu \leq \frac{1}{4^ba}$ then it is not possible to have $e_b>e_\text{thres}$ since it would mean that $\lambda_b=0$ and so $\nu = \frac{1}{4^ba}e^{e_ba} \geq \frac{1}{4^ba}$ which is in contradiction with the hypothesis. Therefore, if  $\nu \leq \frac{1}{4^ba}$ then $e_b = e_\text{thres}$.

If $\nu > \frac{1}{4^ba}$ then by the same logic as before $e_b>e_\text{thres}$. In this case, $\lambda_b = 0$ and so $e_b = \frac{1}{a}\log(\nu 4^b a)$.

\bibliography{Journal.bib}

\begin{thebibliography}{10}
\providecommand{\url}[1]{#1}
\csname url@samestyle\endcsname
\providecommand{\newblock}{\relax}
\providecommand{\bibinfo}[2]{#2}
\providecommand{\BIBentrySTDinterwordspacing}{\spaceskip=0pt\relax}
\providecommand{\BIBentryALTinterwordstretchfactor}{4}
\providecommand{\BIBentryALTinterwordspacing}{\spaceskip=\fontdimen2\font plus
\BIBentryALTinterwordstretchfactor\fontdimen3\font minus
  \fontdimen4\font\relax}
\providecommand{\BIBforeignlanguage}[2]{{%
\expandafter\ifx\csname l@#1\endcsname\relax
\typeout{** WARNING: IEEEtran.bst: No hyphenation pattern has been}%
\typeout{** loaded for the language `#1'. Using the pattern for}%
\typeout{** the default language instead.}%
\else
\language=\csname l@#1\endcsname
\fi
#2}}
\providecommand{\BIBdecl}{\relax}
\BIBdecl

\bibitem{kalman_new_1960}
R.~E. Kalman, ``A {New} {Approach} to {Linear} {Filtering} and {Prediction}
  {Problems},'' \emph{Journal of Basic Engineering}, vol.~82, no.~1, pp.
  35--45, Mar. 1960.

\bibitem{lai_iot_2019}
X.~Lai, T.~Yang, Z.~Wang, and P.~Chen, ``\BIBforeignlanguage{en}{{IoT}
  {Implementation} of {Kalman} {Filter} to {Improve} {Accuracy} of {Air}
  {Quality} {Monitoring} and {Prediction}},''
  \emph{\BIBforeignlanguage{en}{Applied Sciences}}, vol.~9, no.~9, p. 1831,
  Jan. 2019.

\bibitem{anania_development_2008}
G.~Anania, A.~Tognetti, N.~Carbonaro, M.~Tesconi, F.~Cutolo, G.~Zupone, and
  D.~D. Rossi, ``Development of a novel algorithm for human fall detection
  using wearable sensors,'' in \emph{{IEEE} {SENSORS}}, Oct. 2008, pp.
  1336--1339.

\bibitem{sung_simplified_2020}
K.~Sung and H.~Kim, ``Simplified {KF}-based energy-efficient vehicle
  positioning for smartphones,'' \emph{Journal of Communications and Networks},
  vol.~22, no.~2, pp. 93--107, Apr. 2020.

\bibitem{horowitz_11_2014}
M.~Horowitz, ``1.1 {Computing}'s energy problem (and what we can do about
  it),'' in \emph{{IEEE} {International} {Solid}-{State} {Circuits}
  {Conference} {Digest} of {Technical} {Papers} ({ISSCC})}, Feb. 2014, pp.
  10--14.

\bibitem{dreslinski_near-threshold_2010}
R.~G. Dreslinski, M.~Wieckowski, D.~Blaauw, D.~Sylvester, and T.~Mudge,
  ``Near-{Threshold} {Computing}: {Reclaiming} {Moore}'s {Law} {Through}
  {Energy} {Efficient} {Integrated} {Circuits},'' \emph{Proceedings of the
  IEEE}, vol.~98, no.~2, pp. 253--266, Feb. 2010.

\bibitem{kim_generalized_2018}
Y.~Kim, M.~Kang, L.~R. Varshney, and N.~R. Shanbhag, ``Generalized
  {Water}-{Filling} for {Source}-{Aware} {Energy}-{Efficient} {SRAMs},''
  \emph{IEEE Transactions on Communications}, vol.~66, no.~10, pp. 4826--4841,
  2018.

\bibitem{kim_optimizing_2020}
Y.~Kim, Y.~Jeon, C.~Guyot, and Y.~Cassuto, ``Optimizing the {Write} {Fidelity}
  of {MRAMs},'' in \emph{Proceedings of the {IEEE} {International} {Symposium}
  on {Information} {Theory} ({ISIT})}, Jun. 2020, pp. 792--797.

\bibitem{dupraz_binary_2019}
E.~Dupraz and L.~R. Varshney, ``Binary {Recursive} {Estimation} on {Noisy}
  {Hardware},'' in \emph{Proceedings of the {IEEE} {International} {Symposium}
  on {Information} {Theory} ({ISIT})}, 2019, pp. 877--881.

\bibitem{yang_computing_2017}
Y.~Yang, P.~Grover, and S.~Kar, ``Computing {Linear} {Transformations} {With}
  {Unreliable} {Components},'' \emph{IEEE Transactions on Information Theory},
  vol.~63, no.~6, pp. 3729--3756, Jun. 2017.

\bibitem{henwood_layerwise_2020}
S.~Henwood, F.~Leduc-Primeau, and Y.~Savaria, ``Layerwise {Noise}
  {Maximisation} to {Train} {Low}-{Energy} {Deep} {Neural} {Networks},'' in
  \emph{2nd {IEEE} {International} {Conference} on {Artificial} {Intelligence}
  {Circuits} and {Systems} ({AICAS})}, Aug. 2020, pp. 271--275.

\bibitem{hacene_training_2019}
G.~B. Hacene, F.~Leduc-Primeau, A.~B. Soussia, V.~Gripon, and F.~Gagnon,
  ``Training {Modern} {Deep} {Neural} {Networks} for {Memory}-{Fault}
  {Robustness},'' in \emph{{IEEE} {International} {Symposium} on {Circuits} and
  {Systems} ({ISCAS})}, May 2019, pp. 1--5.

\bibitem{yang_fault-tolerant_2016}
Y.~Yang, P.~Grover, and S.~Kar, ``Fault-tolerant distributed logistic
  regression using unreliable components,'' in \emph{54th {Annual} {Allerton}
  {Conference} on {Communication}, {Control}, and {Computing} ({Allerton})},
  Sep. 2016, pp. 940--947.

\bibitem{hegde_energy-efficient_1999}
R.~Hegde and N.~R. Shanbhag, ``Energy-efficient signal processing via
  algorithmic noise-tolerance,'' in \emph{Proceedings. {International}
  {Symposium} on {Low} {Power} {Electronics} and {Design} ({Cat}.
  {No}.{99TH8477})}, Aug. 1999, pp. 30--35.

\bibitem{huang_acoco_2015}
C.~Huang, Y.~Li, and L.~Dolecek, ``{ACOCO}: {Adaptive} {Coding} for
  {Approximate} {Computing} on {Faulty} {Memories},'' \emph{IEEE Transactions
  on Communications}, vol.~63, no.~12, pp. 4615--4628, Dec. 2015.

\bibitem{huang_novel_2018}
Y.~Huang, Y.~Zhang, Z.~Wu, N.~Li, and J.~Chambers, ``A {Novel} {Adaptive}
  {Kalman} {Filter} {With} {Inaccurate} {Process} and {Measurement} {Noise}
  {Covariance} {Matrices},'' \emph{IEEE Transactions on Automatic Control},
  vol.~63, no.~2, pp. 594--601, Feb. 2018.

\bibitem{yang_robust_2001}
G.-H. Yang and J.~L. Wang, ``Robust nonfragile {Kalman} filtering for uncertain
  linear systems with estimator gain uncertainty,'' \emph{IEEE Transactions on
  Automatic Control}, vol.~46, no.~2, pp. 343--348, Feb. 2001.

\bibitem{hounkpevi_robust_2007}
F.~O. Hounkpevi and E.~E. Yaz, ``\BIBforeignlanguage{en}{Robust minimum
  variance linear state estimators for multiple sensors with different failure
  rates},'' \emph{\BIBforeignlanguage{en}{Automatica}}, vol.~43, no.~7, pp.
  1274--1280, Jul. 2007.

\bibitem{nahi_optimal_1969}
N.~Nahi, ``Optimal recursive estimation with uncertain observation,''
  \emph{IEEE Transactions on Information Theory}, vol.~15, no.~4, pp. 457--462,
  Jul. 1969.

\bibitem{jarrah_optimized_2016}
A.~Jarrah, ``Optimized parallel architecture of {Kalman} filter for radar
  tracking applications,'' \emph{Jordan Journal of Electrical Engineering},
  vol.~2, no.~3, pp. 215--230, May 2016.

\bibitem{sunil_kumar_optimization_2020}
T.~Sunil~Kumar and P.~Duraiswamy, ``\BIBforeignlanguage{en}{Optimization of
  {Kalman} {Filter} for {Target} {Tracking} {Applications}},'' in
  \emph{\BIBforeignlanguage{en}{Advances in {Multidisciplinary} {Analysis} and
  {Optimization}}}, ser. Lecture {Notes} in {Mechanical} {Engineering}, R.~R.
  Salagame, P.~Ramu, I.~Narayanaswamy, and D.~K. Saxena, Eds.\hskip 1em plus
  0.5em minus 0.4em\relax Singapore: Springer, 2020, pp. 203--212.

\bibitem{pereira_exploring_2019}
P.~T.~L. Pereira, G.~Paim, P.~Ücker, E.~Costa, S.~Almeida, and S.~Bampi,
  ``Exploring {Architectural} {Solutions} for an {Energy}-{Efficient} {Kalman}
  {Filter} {Gain} {Realization},'' in \emph{26th {IEEE} {International}
  {Conference} on {Electronics}, {Circuits} and {Systems} ({ICECS})}, Nov.
  2019, pp. 650--653.

\bibitem{wang_reducing_2015}
Z.~Wang, J.~Zhang, and N.~Verma, ``Reducing quantization error in low-energy
  {FIR} filter accelerators,'' in \emph{{IEEE} {International} {Conference} on
  {Acoustics}, {Speech} and {Signal} {Processing} ({ICASSP})}, Apr. 2015, pp.
  1032--1036.

\bibitem{xia_energy-efficient_2019}
D.~Xia, Y.~Zhang, P.~Cai, and L.~Huang, ``An {Energy}-{Efficient} {Signal}
  {Detection} {Scheme} for a {Radar}-{Communication} {System} {Based} on the
  {Generalized} {Approximate} {Message}-{Passing} {Algorithm} and
  {Low}-{Precision} {Quantization},'' \emph{IEEE Access}, vol.~7, pp.
  29\,065--29\,075, 2019.

\bibitem{marcastel_energy_2019}
A.~Marcastel, I.~Fijalkow, and L.~Swindlehurst, ``Energy efficient downlink
  massive {MIMO}: {Is} 1-bit quantization a solution ?'' in \emph{16th
  {International} {Symposium} on {Wireless} {Communication} {Systems}
  ({ISWCS})}, Aug. 2019, pp. 198--202.

\bibitem{hashemi_understanding_2017}
S.~Hashemi, N.~Anthony, H.~Tann, R.~I. Bahar, and S.~Reda, ``Understanding the
  impact of precision quantization on the accuracy and energy of neural
  networks,'' in \emph{Proceedings of the {Conference} on {Design},
  {Automation} \& {Test} in {Europe}}, ser. {DATE} '17.\hskip 1em plus 0.5em
  minus 0.4em\relax Leuven, BEL: European Design and Automation Association,
  Mar. 2017, pp. 1478--1483.

\bibitem{ding_quantized_2018}
R.~Ding, Z.~Liu, R.~D.~S. Blanton, and D.~Marculescu, ``Quantized deep neural
  networks for energy efficient hardware-based inference,'' in \emph{23rd
  {Asia} and {South} {Pacific} {Design} {Automation} {Conference}
  ({ASP}-{DAC})}, Jan. 2018, pp. 1--8.

\bibitem{jain_compensated-dnn_2018}
S.~Jain, S.~Venkataramani, V.~Srinivasan, J.~Choi, P.~Chuang, and L.~Chang,
  ``Compensated-{DNN}: energy efficient low-precision deep neural networks by
  compensating quantization errors,'' in \emph{Proceedings of the 55th {Annual}
  {Design} {Automation} {Conference}}, ser. {DAC} '18.\hskip 1em plus 0.5em
  minus 0.4em\relax New York, NY, USA: Association for Computing Machinery,
  Jun. 2018, pp. 1--6.

\bibitem{stripad_performance_1981}
A.~B. Stripad, ``Performance {Degradation} in {Digitally} {Implemented}
  {Kalman} {Filters},'' \emph{IEEE Transactions on Aerospace and Electronic
  Systems}, vol. AES-17, no.~5, pp. 626--634, Sep. 1981.

\bibitem{verhaegen_numerical_1986}
M.~Verhaegen and P.~V. Dooren, ``Numerical aspects of different {Kalman} filter
  implementations,'' \emph{IEEE Transactions on Automatic Control}, vol.~31,
  no.~10, pp. 907--917, 1986.

\bibitem{sun_quantized_2007}
S.~Sun, J.~Lin, L.~Xie, and W.~Xiao, ``Quantized {Kalman} {Filtering},'' in
  \emph{{IEEE} 22nd {International} {Symposium} on {Intelligent} {Control}},
  Oct. 2007, pp. 7--12.

\bibitem{li_distributed_2015}
D.~Li, S.~Kar, and S.~Cui, ``Distributed {Kalman} {Filtering} with quantized
  sensing state,'' in \emph{{IEEE} {International} {Conference} on {Acoustics},
  {Speech} and {Signal} {Processing} ({ICASSP})}, Apr. 2015, pp. 4040--4044.

\bibitem{hu_quantized_2018}
X.~Hu, M.~Bao, X.~Zhang, S.~Wen, X.~Li, and Y.~Hu, ``Quantized {Kalman}
  {Filter} {Tracking} in {Directional} {Sensor} {Networks},'' \emph{IEEE
  Transactions on Mobile Computing}, vol.~17, no.~4, pp. 871--883, Apr. 2018.

\bibitem{you_quantized_2011}
K.~You, L.~Xie, S.~Sun, and W.~Xiao, ``\BIBforeignlanguage{en}{Quantized
  filtering of linear stochastic systems},''
  \emph{\BIBforeignlanguage{en}{Transactions of the Institute of Measurement
  and Control}}, vol.~33, no.~6, pp. 683--698, Aug. 2011.

\bibitem{you_recursive_2009}
K.~You, Y.~Zhao, and L.~Xie, ``Recursive quantized state estimation of
  discrete-time linear stochastic systems,'' in \emph{7th {Asian} {Control}
  {Conference}}, Aug. 2009, pp. 170--175.

\bibitem{dally_digital_2015}
W.~J. Dally, R.~C. Harting, and T.~M. Aamodt, \emph{Digital {Design} {Using}
  {VHDL}: {A} {Systems} {Approach}}.\hskip 1em plus 0.5em minus 0.4em\relax
  Cambridge, United Kingdom: Cambridge University Press, 2015.

\bibitem{kern_improving_2021}
J.~Kern, E.~Dupraz, A.~Aïssa-El-Bey, and F.~Leduc-Primeau, ``Improving the
  {Energy}-{Efficiency} of a {Kalman} {Filter} {Using} {Unreliable}
  {Memories},'' in \emph{{IEEE} {International} {Conference} on {Acoustics},
  {Speech} and {Signal} {Processing} ({ICASSP})}, Jun. 2021, pp. 5345--5349.

\bibitem{ziv_universal_1985}
J.~Ziv, ``On universal quantization,'' \emph{IEEE Transactions on Information
  Theory}, vol.~31, no.~3, pp. 344--347, May 1985.

\bibitem{sripad_necessary_1977}
A.~Sripad and D.~Snyder, ``A necessary and sufficient condition for
  quantization errors to be uniform and white,'' \emph{IEEE Transactions on
  Acoustics, Speech, and Signal Processing}, vol.~25, no.~5, pp. 442--448, Oct.
  1977.

\bibitem{dupraz_analysis_2015}
E.~Dupraz, D.~Declercq, B.~Vasić, and V.~Savin, ``Analysis and {Design} of
  {Finite} {Alphabet} {Iterative} {Decoders} {Robust} to {Faulty} {Hardware},''
  \emph{IEEE Transactions on Communications}, vol.~63, no.~8, pp. 2797--2809,
  Aug. 2015.

\bibitem{ngassa2015density}
C.~K. Ngassa, V.~Savin, E.~Dupraz, and D.~Declercq, ``Density evolution and
  functional threshold for the noisy min-sum decoder,'' \emph{IEEE Transactions
  on Communications}, vol.~63, no.~5, pp. 1497--1509, 2015.

\bibitem{berberidis_data_2017}
D.~Berberidis and G.~B. Giannakis, ``Data {Sketching} for {Large}-{Scale}
  {Kalman} {Filtering},'' \emph{IEEE Transactions on Signal Processing},
  vol.~65, no.~14, pp. 3688--3701, Jul. 2017.

\end{thebibliography}
\bibliographystyle{IEEEtran}

\end{document}